\documentclass[longauth]{aa}  

\usepackage{graphicx}
\usepackage{txfonts}
\usepackage[hidelinks]{hyperref}
\usepackage{amsmath}
\usepackage{comment}
\makeatletter
\renewcommand*\aa@pageof{, page \thepage{} of \pageref*{LastPage}}
\makeatother

\usepackage[separate-uncertainty=true]{siunitx}

\usepackage{natbib}
\bibpunct{(}{)}{;}{a}{}{,}
\begin{document} 

   \title{Polarization analysis of the VLTI and GRAVITY}

   \author{GRAVITY Collaboration\thanks{GRAVITY is developed in a collaboration by the Max Planck Institute for Extraterrestrial Physics, LESIA of Observatoire de Paris/Universit\'e PSL/CNRS /Sorbonne Universit\'e/Universit\'e de Paris and IPAG of Universit\'e Grenoble Alpes / CNRS, the Max Planck Institute for Astronomy, the University of Cologne, the CENTRA - Centro de Astrofisica e Gravita\c c\~ao, and the European Southern Observatory.},
    F.~Widmann\inst{1} 
    \and X.~Haubois\inst{9} 
    \and N.~Schuhler\inst{9} 
    \and O.~Pfuhl\inst{8} 
    \and F.~Eisenhauer\inst{1}
    \and S.~Gillessen\inst{1}
    \and N.~Aimar\inst{2}
    \and A.~Amorim\inst{6, 12}
    \and M.~Baub\"{o}ck\inst{14} 
    \and J.~B.~Berger\inst{5}
    \and H.~Bonnet\inst{8}
    \and G.~Bourdarot\inst{1}
    \and W.~Brandner\inst{3}
    \and Y.~Cl\'{e}net\inst{2} 
    \and R.~Davies\inst{1}
    \and P.~T.~de~Zeeuw\inst{21}
    \and J.~Dexter\inst{13}
    \and A.~Drescher\inst{1} 
    \and A.~Eckart\inst{4,10} 
    \and H.~Feuchtgruber\inst{1}
    \and N.M.~Förster~Schreiber\inst{1} 
    \and P.~Garcia\inst{7, 12}
    \and E.~Gendron\inst{2}
    \and R.~Genzel\inst{1,11}
    \and M.~Hartl\inst{1}
    \and F.~Hau\ss mann\inst{1}
    \and G.~Heißel\inst{19,2}
    \and T.~Henning\inst{3}
    \and S.~Hippler\inst{3} 
    \and M.~Horrobin\inst{4}
    \and A.~Jim\'{e}nez-Rosales\inst{17} 
    \and L.~Jocou\inst{5}
    \and A.~Kaufer\inst{9}
    \and P.~Kervella\inst{2} 
    \and S.~Lacour\inst{2,8}
    \and V.~Lapeyr\`{e}re\inst{2}
    \and J.-B.~Le~Bouquin\inst{5}
    \and P.~L\'{e}na\inst{2}
    \and D.~Lutz\inst{1} 
    \and F.~Mang\inst{1} 
    \and N.~More\inst{1}    
    \and M.~Nowak\inst{15} 
    \and T.~Ott\inst{1} 
    \and T.~Paumard\inst{2} 
    \and K.~Perraut\inst{5} 
    \and G.~Perrin\inst{2} 
    \and S.~Rabien\inst{1} 
    \and D.~Ribeiro\inst{1} 
    \and M.~Sadun~Bordoni\inst{1}
    \and S.~Scheithauer\inst{3} 
    \and J.~Shangguan\inst{1} 
    \and T.~Shimizu\inst{1} 
    \and J.~Stadler\inst{18} 
    \and O.~Straub\inst{20}
    \and C.~Straubmeier\inst{4} 
    \and E.~Sturm\inst{1} 
    \and L.J.~Tacconi\inst{1} 
    \and F.~Vincent\inst{2} 
    \and S.~D.~von~Fellenberg\inst{10} 
    \and E.~Wieprecht\inst{1} 
    \and E.~Wiezorrek\inst{1} 
    \and J.~Woillez\inst{8} 
    }

   \institute{
    Max Planck Institute for Extraterrestrial Physics, Giessenbachstr. 1, D-85748 Garching bei Muenchen, Germany
    \and 
    LESIA, Observatoire de Paris, Universit\'{e} PSL, CNRS, Sorbonne Universit\'{e}, Universit\'{e} de Paris, 5 place Jules Janssen, 92195 Meudon, France 
    \and 
    Max-Planck-Institute for Astronomy, K\"{o}nigsstuhl 17, 69117, Heidelberg, Germany
    \and 
    1. Physikalisches Institut, Universit\"{a}t zu K\"{o}ln, Z\"{u}lpicher Str. 77, 50937, K\"{o}ln, Germany
    \and 
    Univ. Grenoble Alpes, CNRS, IPAG, 38000 Grenoble, France
    \and 
    Universidade de Lisboa - Faculdade de Ci\^{e}ncias, Campo Grande, 1749-016 Lisboa, Portugal
    \and 
    Faculdade de Engenharia, Universidade do Porto, Rua Dr. Roberto Frias, 4200-465 Porto, Portugal
    \and 
    European Southern Observatory, Karl-Schwarzschild-Str. 2, 85748, Garching, Germany
    \and 
    European Southern Observatory, Casilla 19001, Santiago 19, Chile
    \and 
    Max-Planck-Institute for Radio Astronomy, Auf dem H\"{u}gel 69, 53121, Bonn, Germany
    \and 
    Departments of Physics and Astronomy, Le Conte Hall, University of California, Berkeley, CA 94720, USA
    \and 
    CENTRA - Centro de Astrof\'{\i}sica e Gravita\c{c}\~{a}o, IST, Universidade de Lisboa, 1049-001 Lisboa, Portugal
    \and 
    Department of Astrophysical \& Planetary Sciences, JILA, Duane Physics Bldg., 2000 Colorado Ave, University of Colorado, Boulder, CO 80309
    \and  
    Department of Physics, University of Illinois, 1110 West Green Street, Urbana, IL 61801, USA
    \and  
    Institute of Astronomy, University of Cambridge, Madingley Road, Cambridge CB3 0HA, UK
    \and 
    Hamburger Sternwarte, Universit\"{a}t Hamburg, Gojenbergsweg 112, 21029, Hamburg, Germany
    \and 
    Department of Astrophysics, IMAPP, Radboud University, 6500 GL Nĳmegen, The Netherlands
    \and 
    Max Planck Institute for Astrophysics, Karl-Schwarzschild-Str. 1, 85741 Garching, Germany
    \and 
    Advanced Concepts Team, European Space Agency, TEC-SF, ESTEC, Keplerlaan 1, 2201 AZ Noordwijk, The Netherlands
    \and 
    ORIGINS Excellence Cluster, Boltzmannstra\ss e 2, D-85748 Garching bei Muenchen, Germany
    \and
    Leiden University, 2311 EZ Leiden
    }

   \date{Received \today; accepted \today}
  \abstract
   {}
   {The goal of this work is to characterize the polarization effects of the beam path of the Very Large Telescope Interferometer (VLTI) and the GRAVITY beam combiner instrument. This is useful for two reasons: to calibrate polarimetric observations with GRAVITY for instrumental effects and to understand the systematic error introduced to the astrometry due to birefringence when observing targets with a significant intrinsic polarization.}
   {By combining a model of the VLTI light path and its mirrors and dedicated experimental data, we construct a full polarization model of the VLTI Unit Telescopes (UTs) and the GRAVITY instrument. We first characterize all telescopes together to construct a universal UT calibration model for polarized targets with the VLTI. We then expand the model to include the differential birefringence between the UTs. With this, we can constrain the systematic errors and the contrast loss for highly polarized targets.}
   {Together with this paper, we publish a standalone Python package, which can be used to calibrate the instrumental effects on polarimetric observations. This enables the community to use GRAVITY with the UTs to observe targets in a polarimetric observing mode. We demonstrate the calibration model with the galactic center star IRS~16C. For this source, we can constrain the polarization degree to within \SI{0.4}{\percent} and the polarization angle within \SI{5}{\degree} while being consistent with the literature values. Furthermore, we show that there is no significant contrast loss, even if the science and fringe-tracker targets have significantly different polarization, and we determine that the phase error in such an observation is smaller than \SI{1}{\degree}, corresponding to an astrometric error of \SI{10}{\micro as}.}
   {With this work, we enable the use of the polarimetric mode with GRAVITY/UTs for the community and outline the steps necessary to observe and calibrate polarized targets with GRAVITY. We demonstrate that it is possible to measure the intrinsic polarization of astrophysical sources with high precision and that polarization effects do not limit astrometric observations of polarized targets.}

\maketitle

\nolinenumbers

\section{Introduction}
Polarization is an essential part of the information contained in electromagnetic radiation of astronomical sources. The use of polarimetric observations enables a better understanding of the source of radiation as well as its environment. Polarimetric observations are nowadays used over a very broad range of science cases, and an increasing number of instruments are equipped with a polarimetric mode \citep[see e.g.][]{Witzel2011, Dorn2014, Norris2015, vanHolstein2020}. With the enormous success of the Very Large Telescope Interferometer (VLTI) beam combiner instrument GRAVITY over the last years in various science fields \citep[see e.g.][]{gravity2018redshift, GRAVITY2018, Gravity2018AGN, Gravity2019exo, Gravity2019YSO}, interest in polarimetric observations with GRAVITY and the VLTI has also grown. The fundamental capabilities of GRAVITY to make polarimetric observations have already been shown by observing the polarization of flares from the supermassive black hole SgrA* \citep{GRAVITY2018, Gravity2020pol}. With the help of the polarization data, it was possible to constrain the magnetic fields around SgrA*. Similarly, the EHT collaboration has studied the magnetic fields around the black hole M87* with their recently released polarimetric image \citep{EHT2021}. But not only the study of magnetic fields is enabled by polarimetry, but also many other research areas profit from the availability of polarization measurements. For example, disks around young stellar objects can be studied with the help of polarimetry \citep{Hunziker2021}, or measurements of dust properties of evolved stars benefit from polarization measurements \citep{Ireland2005, Norris2012b, Haubois2019}. For a more complete overview of polarized observations, see \cite{Elias2008} and \cite{Trippe2014}. To combine polarimetric measurements with the unique angular resolution of GRAVITY, we want to characterize the polarization properties of GRAVITY  and the VLTI.

Ideally, a telescope and its instrument would not alter the polarization of incoming light. In reality, however, the optical train of a telescope influences the polarization signal. This can mean that the instrument produces a polarization signal (so-called instrumental polarization (IP)) or alters the incoming polarization by introducing crosstalk, which mixes the incoming polarization states. To compensate for these effects, the telescope and its instrument must be carefully calibrated for their effect on the measured polarization signal. In this paper, we show the results of a series of measurements to calibrate the polarimetric properties of the VLTI. This includes characterizing the amount of crosstalk between different polarization states and the IP introduced by the VLTI. In the case of an interferometer, this is more difficult than for a single-telescope instrument, as there are a significantly higher number of reflections, and one also has to account for both of the rotations of the telescope, in elevation as well as in azimuth. For this reason, polarimetric observations with optical interferometers are not common yet, but the foundations were laid in the early 2000s \citep{Elias2001, Elias2004}. The first steps were done soon after \citep{Ireland2005, Perraut2006} to study variable stars and circumstellar environments, and later in aperture masking \citep{Norris2012, Norris2015}. Today, similar to GRAVITY at the VLTI, MIRC-X at CHARA has its first polarimetric observations \citep{Setterholm2020}. 

 While the modeling is more complicated for an interferometer than for a single telescope, there is no fundamental difference in the calibration model for absolute polarization.  We can use similar calibration models to those used in solar physics \citep[see e.g.][]{Beck2005, Harrington2019} and for the NACO and SPHERE instruments at the VLT \citep{Witzel2011, vanHolstein2020}. We then use our test data to adapt the model and constrain the polarimetric properties of the VLTI. With this approach, we construct a full calibration model to correct polarimetric observations. 

Apart from the absolute effect the VLTI has on the polarization measurement, there is an additional effect that has to be considered for interferometers. If the light paths of the individual telescopes have different polarimetric properties, this can introduce differential birefringence between the different telescopes. The VLTI has been built with great care to make sure that the different light paths and reflections within are as similar as possible, but of course, this cannot be ideal, as there are imperfections in the trains, as well as individual upgrades such as the adaptive secondary mirror at UT4. These differential effects are important to understand, as differential birefringence leads to a loss of fringe contrast and therefore limits the sensitivity of an interferometer \citep{Beckers1990, Perraut1996}. Furthermore, differential effects can also introduce errors to the visibility phase, limiting the astrometric accuracy for polarized targets. This was already explored for a part of the VLTI by \cite{Lazareff2014} and is continued with this work.

While most of the light path and the reflections are similar for the Unit Telescopes (UTs) and the Auxiliary Telescopes (ATs), we focus solely on UT observations in this work. For the ATs, the derotation of the field is done in the telescopes themselves, which adds more complexity to the polarization measurement. Furthermore, the ATs are not fixed in place but can be repositioned. This could affect the polarization, mainly if the telescopes are located on different sides of the delay line. Considering this and the scientific importance of the UTs, we decided to limit this study to the UTs.

The work presented here is split into two main parts. In the first part (\autoref{sec:convention_st} - \ref{sec:data}), we develop a calibration model for the VLTI and GRAVITY to calibrate polarimetric observations with GRAVITY. For this part, we assume that all telescopes are identical, and we use the Stokes formalism. This is the formalism typically used for modeling instrumental effects, and it minimizes the necessary degrees of freedom in the model. This formalism is introduced in \autoref{sec:convention_st} before we discuss the instrumental effects of the VLTI on the polarization with the telescope model and the calibration measurements in sections \ref{sec:model} and \ref{sec:calibration}. In \autoref{sec:gravity}, we will add GRAVITY to the model to have a complete model and apply it to on-sky data in \autoref{sec:full_calib} - \ref{sec:data}. 

In the second part of the work (\autoref{sec:differential}), we investigate differential effects between the telescopes and how they affect observations. For this part, we have to analyze the data of each telescope individually to measure differential effects between them. We also switch to the Jones formalism, which needs more parameters to describe the polarization but can describe the propagation of the light phase through the telescopes, which is needed to understand the effects of differential behavior on the interferometric signal. We, therefore, introduce the Jones formalism and the necessary concepts in \autoref{sec:jones}. Similar to the first part, we then fit the model again, but this time with the Jones formalism and for each telescope individually to be able to constrain differential effects. The results of both parts are shortly summarized in \autoref{sec:conclusion}.

\section{Conventions - Stokes formalism}\label{sec:convention_st}
There are two different conventions for describing polarization \citep{Collett1992, Tinbergen2005}. One is the Stokes formalism (with Stokes vectors and Mueller matrices), and the other is the Jones formalism  (with Jones vectors and Jones matrices). The Stokes formalism is often used to describe instrumental effects on polarization, as the components of the Stokes vector directly relate to the measurable intensities. It can also describe partial polarization and has simple formulas to measure and calculate the fundamental properties of polarized light. The Stokes values are also easily measured using a half- and a quarter-wave plate. One disadvantage is that the Stokes formalism does not include phase information, which we need to describe interferometric quantities.
We will, therefore, start with the Stokes formalism and later switch to the Jones formalism when the phase information is needed. This is the case when we look at differential effects between two telescopes in \autoref{sec:differential}. 

In the Stokes formalism, the light and its polarization are described by a Stokes vector. An electric field which is described by
\begin{equation}
        \vec{E}(z,t) = \left(\begin{array}{c} 
                            E_x \\ 
                            E_y
                        \end{array}\right)e^{i(kz-\omega t)}
               = \left(\begin{array}{c} 
                A_x \cdot e^{i\phi_x}\\ 
                A_y \cdot e^{i\phi_y}
        \end{array}\right)e^{i(kz-\omega t)},
\end{equation}
the Stokes vector is defined as
\begin{equation}
    s = \left[\begin{array}{c}
                    I\\
                    q\\
                    u\\
                    v
                \end{array}\right]
    = \left[\begin{array}{c}
                    \langle E_x^2 +E_y^2 \rangle\\
                    \langle E_x^2 -E_y^2 \rangle\\
                    \langle 2E_xE_y \cos\delta\rangle\\
                    \langle 2E_xE_y \sin\delta\rangle
                \end{array}\right]
    = \left[\begin{array}{c}
                    A_x^2 +A_y^2\\
                    A_x^2 -A_y^2\\
                    2A_xA_y \cos\delta\\
                    2A_xA_y \sin\delta
                \end{array}\right],
\end{equation}
with $\delta=\phi_x-\phi_y$. 
For practical purposes, it is easier to define the Stokes vector with the measured flux at different angles:
\begin{equation}
\label{equ:stokes_measure}
    s = \left[\begin{array}{c}
                    I\\
                    q\\
                    u\\
                    v
                \end{array}\right]
    = \left[\begin{array}{c}
                    F_{00} + F_{90} \\
                    F_{00} - F_{90} \\
                    F_{45} - F_{135} \\
                    F_{RH} - F_{LH} \\
                \end{array}\right],
\end{equation}
where $F_{00}$ is the flux after a linear polarization filter at \SI{0}{\degree} and $F_{RH}$ and $_{LH}$ are the flux measurement for right- and left-handed circular polarization. As the absolute intensity is not important for the polarization properties, we will in the following only consider normalized Stokes vectors:
\begin{equation}
    S = \frac{s}{I} = \left[\begin{array}{c}
                    1\\
                    Q\\
                    U\\
                    V
                \end{array}\right].
\end{equation}
The first parameter is the intensity in the non-normalized Stokes vector and is one in the normalized Stokes vector. The second and third parameters, Q and U, represent linear polarization. Positive Q shows linear polarization in the vertical direction and negative Q in the horizontal direction. U is \SI{45}{\degree} rotated in \textbf{counterclockwise direction} with respect to Q, looking towards the source. V describes circular polarization with positive V being \textbf{right-handed} and negative V being \textbf{left-handed}\footnote{These definitions follow the IAU recommendations; see Transactions of the IAU, Vol. XVB, pg. 166.}. The normalized Stokes vectors Q, U, and V range from -1 to 1.

A 4x4 real matrix, the Mueller matrix, describes the polarization change for any optical system. For an input state $S_{in}$ the output state $S_{out}$ is calculated as follows:
\begin{equation}\label{equ:smueller}
    S_{out}=M\cdot S_{in}.
\end{equation}
The general components of the Mueller matrix are described as follows:
\begin{equation}
M = \left( 
\begin{array}{cccc}
I \rightarrow I & Q \rightarrow I & U \rightarrow I & V \rightarrow I \\
I \rightarrow Q & Q \rightarrow Q & U \rightarrow Q & V \rightarrow Q \\
I \rightarrow U & Q \rightarrow U & U \rightarrow U & V \rightarrow U \\
I \rightarrow V & Q \rightarrow V & U \rightarrow V & V \rightarrow V \\
\end{array}.
\right)
\label{equ:schem_mueller}
\end{equation}
Regarding instrumental effects, we look at two main contributions in the Mueller matrix. The first one is induced polarization, often referred to as instrumental polarization (IP). This is described by the first column of the Mueller matrix: $I\rightarrow Q/U/V$ and corresponds to the polarization signal produced by the instrument. The second effect is the crosstalk between the states $Q/U/V \leftrightarrow U/V/Q$. The crosstalk introduces a mixing of the polarization states by the instrument. The first row of the Mueller matrix is often considered not important in astrophysical implications, as the elements $Q/U/V \rightarrow I$ do not play a significant role as, for most cases, the source polarization is small ($Q, U, V \ll 1$). The quantities on the diagonal are the element $I \rightarrow  I$, which is one when we work with normalized Stokes vectors and the terms for polarimetric efficiency $Q/U/V \leftrightarrow Q/U/V$, which describes how well polarimetric states are maintained.

One of the advantages of the Stokes parametrization is that it is very easy to calculate the essential polarization properties. From the Stokes parameters, one can calculate the degree of polarization (DOP), the degree of linear polarization (DOLP), and the polarization angle ($\Theta_{\mbox{pol}}$) as follows:
\begin{equation}
    \mbox{DOP} = \sqrt{Q^2 + U^2 + V^2},
\end{equation}
\begin{equation}
    \mbox{DOLP} = \sqrt{Q^2 + U^2},
\end{equation}
\begin{equation}
    \Theta_{\mbox{pol}} = \frac{1}{2} \arctan\left(\frac{U}{Q}\right) + n\cdot\textstyle\frac{\pi}{2},
\end{equation}
where $n$ is 1 for Q $<$ 0 and otherwise 0.

\section{VLTI model}\label{sec:model}
\begin{figure}
	\centering
    \includegraphics[width=0.45\textwidth]{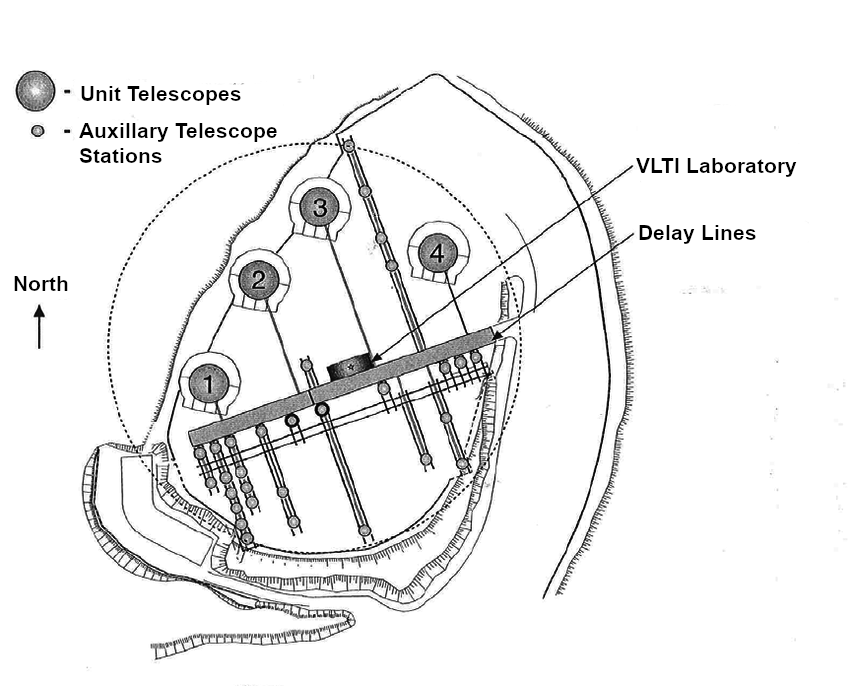}
	\caption{Schematic layout of the VLTI platform. The position of the four UTs is shown in big circles, and the possible stations for ATs in small circles. The delay lines and the VLTI Lab are indicated in the center of the platform.}
	\label{fig:sketch_platform}
\end{figure}
\begin{figure*}[t]
	\centering
    \includegraphics[width=0.95\textwidth]{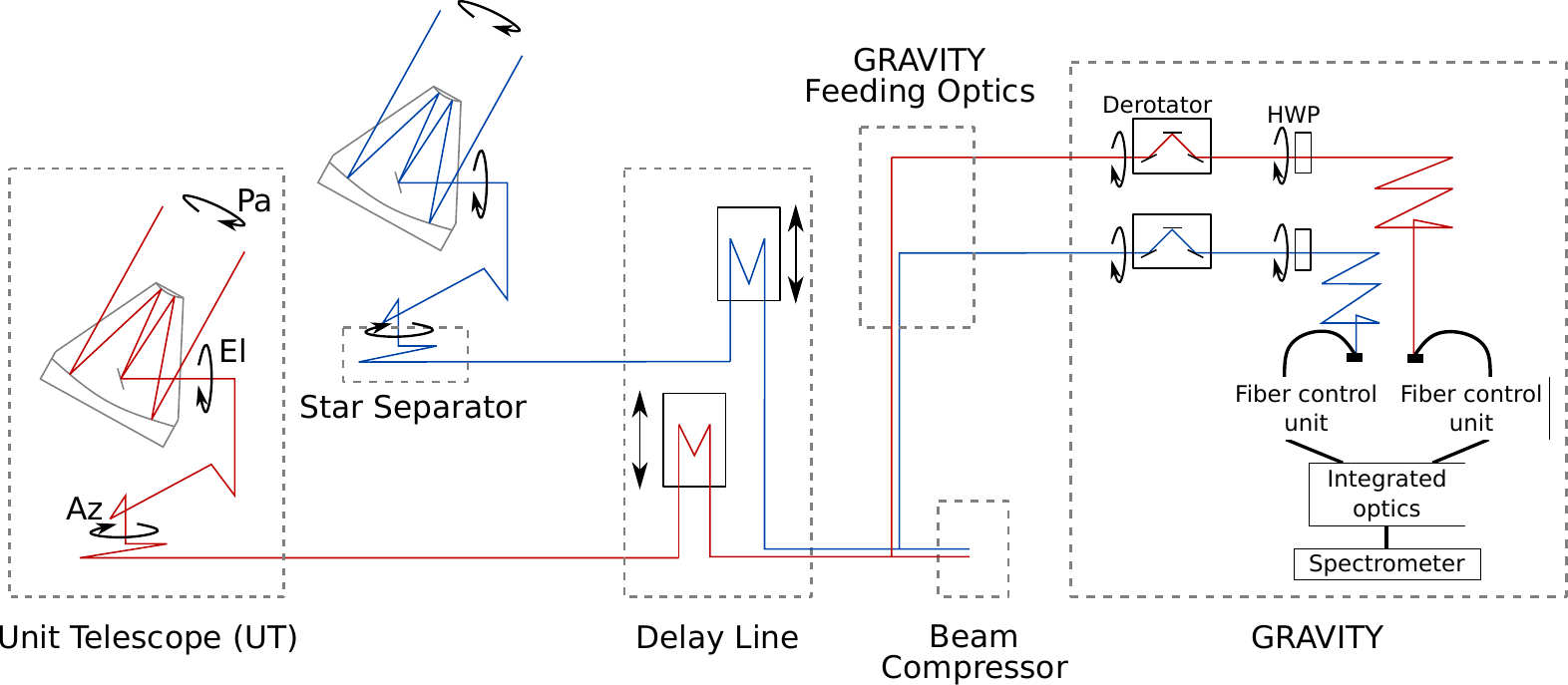}
	\caption{Schematic model of the VLTI light path. The light path is shown for two UTs in red and blue, including GRAVITY. The rounded arrows indicate the possible rotations in the VLTI (for change in Azimuth (Az), Elevation (El), and paralactic angle (Pa)) and in GRAVITY (at the field derotator and the half-wave plate (HWP)). The straight arrows indicate the movement of the delay lines. The location of the star separator is only shown for the blue beam but is in the same place for both telescopes. The fiber optics and elements within the GRAVITY beam combiner are only indicated here and further explained in \autoref{sec:gravity}.}
	\label{fig:sketch_path}
\end{figure*}

To build up a calibration model for the VLTI, we first model the light path with all its mirrors. This is done only once and not for all UTs individually, as the light paths of the four UTs are almost identical. The only differences are the distances between some mirrors and the directions of the first reflection in the delay line. This reflection changes depending on whether the telescope beam is on the left or right of the VLTI lab, so it is different between UT 1 \& 2 and UT 3 \& 4 (see \autoref{fig:sketch_platform}). As this changes the incident plane by \SI{180}{\degree}, it does not affect the propagation of polarization. The overall model is based on what was previously developed by \cite{Lazareff2014} \citep[for more details, see also:][]{Lazareff2014b}.

\subsection{VLTI light path}
The light path for two telescopes is shown in \autoref{fig:sketch_path} \& \ref{fig:sketch_exp}. After the primary and secondary mirrors, the light is sent to the Nasmyth platform by M3. It then travels to the center below the telescopes, where it is guided into the Coud\'{e} room. In the Coud\'{e} room, it travels through the star separator, where some of the light is directed into the adaptive optics system. As the adaptive optics system is not sensitive to polarization and the light is not fed back, it is irrelevant for studying the polarization properties. From there, all light beams are sent to the delay lines. Here, the positions of the mirrors change slightly for each telescope, but as the distances only differ in the direction of light propagation and all the reflections are identical, these differences do not affect the polarization. From the delay lines, the light enters the VLTI lab, reaching the beam compressor, which adapts the beam size to fit the beam size required by GRAVITY. After the beam compressor, the light continues to the VLTI switchyard, where it can be sent to the individual instruments. For GRAVITY, there is one more reflection to feed the light into the instrument. In the instrument, the field is derotated by a K-Mirror and then passes a half-wave plate (HWP) before it is fed into the fiber coupler. More details on the exact components of GRAVITY are given in \autoref{sec:gravity}.

\subsection{Modelling}\label{sec:modeling}
An electromagnetic wave incident on a mirror can be decomposed into a component parallel (p-component) and a component perpendicular (s-component) to the plane of incidence. Reflections on a metallic mirror can introduce a linear polarization if the reflectivity of the two components is different or a circular polarization when there is a different phase shift for the two components. The Mueller matrix which describes such a reflection is given by \citep[see e.g.][]{Collett1992}:
\begin{equation}
\label{equ:mueller}
M = \frac{1}{2}\left( 
\begin{array}{cccc}
r_s^2+r_p^2 & r_s^2-r_p^2 & 0 & 0 \\
r_s^2-r_p^2 & r_s^2+r_p^2 & 0 & 0 \\
0 & 0 & 2r_sr_p\cos(\delta) & 2r_sr_p\sin(\delta) \\
0 & 0 & -2r_sr_p\sin(\delta) & 2r_sr_p\cos(\delta) 
\end{array}
\right)
\end{equation}
where $r$ is the reflection coefficient of each component and $\delta$ the relative retardation: $\delta = \phi_s-\phi_p$. $r$ and $\delta$ can be directly calculated from the Fresnel formula:
\begin{equation}
\sin\Theta_i = n \sin\Theta_t,
\end{equation}
where $\Theta_i$ and $\Theta_t$ are the angles of incident and transmitted light and $n$ is the material and wavelength dependent refractive index. While this is the original Fresnel formula, the refractive index for metals is a complex number, and therefore, the reflection angle $\Theta_t$ is complex and is not a regular angle anymore. With the incident angle and the complex $\Theta_t$, one can now  calculate the reflectance:
\begin{equation}\label{equ:fresnels}
R_s = -\frac{\sin(\Theta_i-\Theta_t)}{\sin(\Theta_i+\Theta_t)} = r_s\exp(i\phi_s),
\end{equation}
\begin{equation}\label{equ:fresnelp}
R_p = \frac{\tan(\Theta_i-\Theta_t)}{\tan(\Theta_i+\Theta_t)} = r_p\exp(i\phi_p),
\end{equation}
While the transmitted part is not relevant for metal surfaces, one can still use n and $\Theta_i$ to calculate the reflection coefficients and formulate the Mueller matrix for the reflection of a mirror.

As the form of the Mueller matrix given in \autoref{equ:mueller} is rather unintuitive, one can modify it by introducing the diattenuation $D$:
\begin{equation}
    D = \frac{r_s^2-r_p^2}{r_s^2+r_p^2}. 
\end{equation}
With this, the Mueller matrix of a metal surface takes the form \citep{Chenault1993, Keller2002, vanHolstein2020}:
\begin{equation}
\label{equ:mueller_mod}
M = \frac{r_s^2+r_p^2}{2}\left( 
\begin{array}{cccc}
1 & D  & 0 & 0 \\
D & 1 & 0 & 0 \\
0 & 0 & \sqrt{1-D^2}\cos(\delta) & \sqrt{1-D^2}\sin(\delta) \\
0 & 0 & -\sqrt{1-D^2}\sin(\delta) & \sqrt{1-D^2}\cos(\delta) 
\end{array}
\right).
\end{equation}
This form has the advantage that one can disentangle the different effects of a single mirror: The diattenuation shows the amount of instrumental polarization. It has values between $-1$ and $1$ and is $0$ for no instrumental polarization. The relative retardation, or retardance, $\delta$ introduces crosstalk for values below \SI{180}{\degree}. The prefactor to the Mueller matrix in \autoref{equ:mueller_mod} is only important for the total transmission and cancels out when working with normalized Stokes vectors. For the later calculations, we will set it to one.

\begin{figure*}[t]
	\centering
    \includegraphics[width=0.55\textwidth]{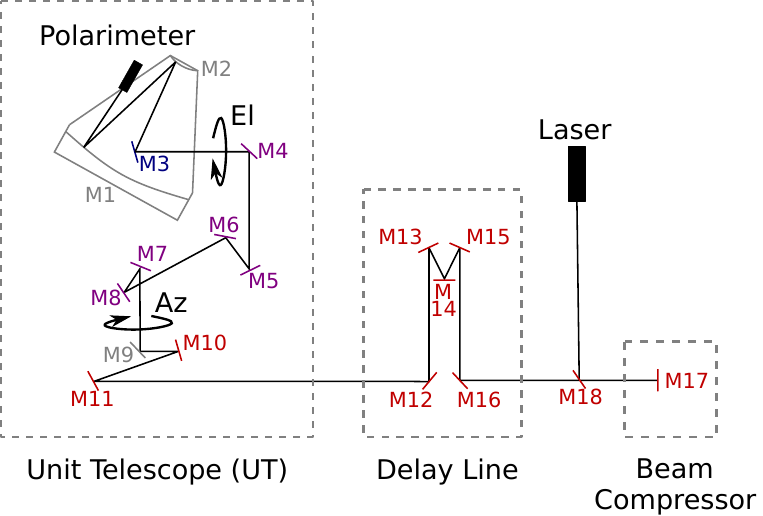}
	\caption{Simplified version of the VLTI light path from \autoref{fig:sketch_path} to show the modeling and the experimental setup. The black rectangles show where the laser is launched and where the polarimeter is mounted. The names of the mirrors used in the text are given. The color of the mirror number shows the grouping which was used for the fitting. Grey mirrors are not fitted in our calibration model. Each colored group is located in one common plane: M4-M8 are in one vertical plane, and M10 - M18 are in one horizontal plane.}
	\label{fig:sketch_exp}
\end{figure*}
To get the incidence angles of the mirrors, we use the positions of the individual mirrors as given by \cite{Michel2000}. We show the notation for the mirrors, which we use in the following, in \autoref{fig:sketch_exp}. From the positions, we can calculate the light path and the incident angle at each mirror. Together with the material of the mirrors, this is enough to set up the VLTI model. However, we implemented the following simplifications:
\begin{itemize}
	\item We do not model M1 and M2 as the incidences are near normal, and we can neglect their contribution.
	\item The star separator (see \autoref{fig:sketch_path}) is not implemented as a special element in our model. As it is not rotating, we instead approximated it with the initial positions of M10 and M11 from the VLTI setup before the implementation of the star separator, as given by \cite{Michel2000} (see also \autoref{fig:sketch_exp}).
	\item The reflection in the delay line is done with a cat-eye retroreflector. In our model, we simplify this to three mirrors.
	\item The beam compressor is modeled as just one equivalent mirror, as all the incidences are very close to normal.
\end{itemize}

The materials for the individual mirrors and the incidence angles are listed in \autoref{tab:mirrors_model}. The mirror M9 is a dichroic mirror, which reflects the infrared light and passes the optical light to the optical adaptive optics system. As this is not a simple metal surface, the polarimetric quantities for this mirror were measured by \cite{Lazareff2014}. The mirrors in the cat-eye are modeled as gold mirrors, but as mentioned before, this is just an approximation as the cat-eye retroreflector should not have a significant influence as all reflections are near normal. One thing to consider is that the silver mirrors in the train have a protective coating, which will lead to a different effect on polarized light.

The refractive indices of the mirrors are taken from the initial model from \cite{Lazareff2014}, as given in an online database \footnote{\url{https://refractiveindex.info/}}. For the three used mirror materials the refractive indices at a wavelength of \SI{2250}{\nano\meter} are:
\begin{itemize}
	\item Gold: $\text{n} = 0.99 + 13.81i$ 
	\item Silver: $\text{n} = 0.77 + 13.41i$
	\item Aluminum: $\text{n} = 2.75 + 22.28i$
\end{itemize}
GRAVITY operates in the K-Band between 2000 and \SI{2500}{\nano\meter}, which is why we use the refractive indices at the center of this band, at \SI{2250}{\nano\meter}. The differences in refractive indices over the $\pm$ \SI{250}{\nano\meter} are on the order of $\pm (0.1 + 2i)$ and should not have a significant effect on the calibration model we are building up. With the entire model, we later analyze the wavelength dependency more, which is then discussed in \autoref{app:model}.

For the gold and aluminum, we can directly use these values, but for the silver mirrors, we have to consider that they have a protective coating. For these mirrors, we assume a protective layer of $Al_2O_3$ with a thickness of \SI{210}{\nano\meter}. These values fit the measurement we have available for one of the protected mirrors. The diattenuation and the phase shift can still describe the polarimetric properties of the protected mirrors, but the calculation is more complicated than outlined before. We use the method given by \cite{Jellison1999, Goldstein2003}. For the used method and a complete treatment of the protected mirrors, see \autoref{app:protected}.

\begin{table}
\caption{Material and incidence angle of the mirrors used for the model of the VLTI.}             
\label{tab:mirrors_model}
\centering                          
\begin{tabular}{r c c}       
\hline\hline         
\rule{0pt}{2ex} \rule{0pt}{2ex} & Material & Incidence angle [$^\circ$] \\
\hline 
\rule{0pt}{2ex} M3 \rule{0pt}{2ex}  & Aluminium  &  45\\
\rule{0pt}{2ex} M4 \rule{0pt}{2ex}  & Silver  &  45\\
\rule{0pt}{2ex} M5 \rule{0pt}{2ex}  & Silver  &  4\\
\rule{0pt}{2ex} M6 \rule{0pt}{2ex}  & Silver  &  25\\
\rule{0pt}{2ex} M7 \rule{0pt}{2ex}  & Silver  &  7\\
\rule{0pt}{2ex} M8 \rule{0pt}{2ex}  & Silver  &  13\\
\rule{0pt}{2ex} M9 \rule{0pt}{2ex}  & Dichroic Mirror  &  45\\
\rule{0pt}{2ex} M10 \rule{0pt}{2ex}  & Silver  &  5\\
\rule{0pt}{2ex} M11 \rule{0pt}{2ex}  & Silver  &  2\\
\rule{0pt}{2ex} M12 \rule{0pt}{2ex}  & Silver  &  45\\
\rule{0pt}{2ex} M13 \rule{0pt}{2ex}  & Cat-eye, Gold  &  5\\
\rule{0pt}{2ex} M14 \rule{0pt}{2ex}  & Cat-eye, Gold  &  11\\
\rule{0pt}{2ex} M15 \rule{0pt}{2ex}  & Cat-eye, Gold  &  5\\
\rule{0pt}{2ex} M16 \rule{0pt}{2ex}  & Silver  &  45\\
\rule{0pt}{2ex} M17 \rule{0pt}{2ex}  & Gold  &  0\\
\rule{0pt}{2ex} M18 \rule{0pt}{2ex}  & Gold &  45\\
\hline       
\end{tabular}
\end{table}

With the phase shift and diattenuation for each mirror, we can calculate the Mueller matrix for each mirror. To combine several Mueller matrices, one can take the product of them to get the combined Mueller matrix:
\begin{equation}
M = M_nM_{n-1} \cdots M_2M_1.
\end{equation}
Our model starts at M3 and goes all the way down into the VLTI lab until the GRAVITY feeding optics.

\subsection{Coordinate system \& Field rotation}\label{sec:fieldrot}
To finalize the polarization model, one has to consider the 3D nature of the light path and that there are several fixed and varying field rotations in the path of the VLTI. The initial coordinate system is chosen so that Q aligns with North. From this start, the model is constructed by using the mirror positions from \cite{Michel2000} and defining the light path as the vector from one mirror to the next. For each mirror, we calculate the angle of incidence and use this to define the direction of the s- and p-component before and after the mirror. For each pair of consecutive mirrors, we then compare the direction of the s- \& p-components of the outgoing light with the same components of the incoming light of the next mirror. If there is a field rotation between the two mirrors, the comparison shows exactly this rotation, which we add to the model. In this way, we follow a plane through the VLTI that is orthogonal to the direction of light propagation. With this approach, we get the direction of s- \& p- components and all field rotations. This model of following the propagating light while allowing for rotations in between does fully describe the light path, and no additional assumptions have to be made \citep[For examples of similar approaches see e.g.][]{Capitani1989, Beck2005, Balthasar2011, Harrington2019}.

As the analysis is done for a telescope at the reference position (azimuth at \SI{0}{\degree} and elevation at \SI{90}{\degree}), we have to add the field rotation for azimuth and elevation by hand. The resulting field rotations are the following (see also \autoref{fig:sketch_exp}):
\begin{itemize}
    \item Between M3 and M4, there is a rotation due to the telescope movement in elevation (El). The rotation is $z = 90^\circ - \mbox{El}$, as an elevation of $0^\circ$ corresponds to a zenith angle of $90^\circ$.
    \item Between M8 and M9, there is a rotation depending on the telescope's position in azimuth (Az). The rotation is given by $\phi = -(\mbox{Az} + 18.98^\circ) + 6.02^\circ$. The $18.98^\circ$ comes from the fact that the VLTI baselines are rotated by $-18.98^\circ$ compared to the east-west direction (see \autoref{fig:sketch_platform}).  As the zero position of the UTs is towards the south, this introduces an offset in the azimuth position. The $6.02^\circ$ comes from a field rotation in the star separator mirrors.
    \item Between M9 and M10, there is a \SI{90}{\degree} rotation as the plane of reflection changes from reflections perpendicular to the ground (in the telescopes) to reflections in a horizontal plane (in the delay lines and the VLTI lab)
    \item One additional rotation, which is not coming out of the model, is the paralactic angle. This has to be taken into account as the Stokes parameters are defined towards the source and not in our chosen reference system (towards the north)
\end{itemize}
All these rotations are identical for the light paths of the four UTs. This leads to a total rotation of the field in the light path by:
\begin{equation}
\label{equ:fieldrot}
\begin{aligned}
    \Theta_{VLTI} & = (90^\circ - \mbox{El}) - (\mbox{Az} + 18.98^\circ) + 6.02^\circ + 90^\circ + \mbox{Pa}\\
    & = \mbox{Pa} - \mbox{El} - \mbox{Az} + 167.04^\circ.
\end{aligned}
\end{equation}
This is also stated in \cite{Gitton2003}, with the only difference being the sign of the azimuth angle. This is because the angle is defined as East of South in \cite{Gitton2003}, while we use the convention of the ESO ISS system, which is East of North \citep{Perraut2010}. 

A rotated optical element would usually be implemented by multiplying the Mueller matrix of the element with a rotation matrix R. The Mueller matrix of that element is given by $M_\Theta = R(-\Theta)\cdot M \cdot R(\Theta)$, where R is \citep{Collett1992}:
\begin{equation}
R(\Theta) = \left( 
\begin{array}{cccc}
1 & 0 & 0 & 0 \\
0 & \cos{2\Theta} & \sin{2\Theta} & 0 \\
0 & -\sin{2\Theta} & \cos{2\Theta} & 0 \\
0 & 0 & 0 & 1 
\end{array}
\right).
\label{equ:rotmat}
\end{equation}
This is done to ensure that the input coordinate system is preserved. In our case, the field rotations are part of the optical system, and we have no advantage of preserving the input coordinate system. We, therefore, rotate the coordinate system at each field rotation. The response of an optical element and the field rotation is described by $S = R(\Theta) \cdot ( M \cdot S') = (R(\Theta) \cdot  M) \cdot S'$. This way, the field rotation as given in \autoref{equ:fieldrot} is automatically included in the final calibration Mueller matrix, with the resulting reference system then being defined as Q vertical in the lab. This approach gives the same result as using the usual convention that conserves the coordinate system and then rotating the reference system by the full angle given in \autoref{equ:fieldrot}.

One must take into account that each metallic mirror introduces a \SI{180}{\degree} phase shift, which is equivalent to a change of coordinate system for a Stokes vector \citep{Keller2002}. Due to this change in the coordinate system, rotations after an odd number of mirrors go into the total field rotation in the opposite direction. 
The full Mueller matrix of the VLTI, including all necessary rotations, is then given by:
\begin{equation}
\begin{aligned}
    M_{VLTI} = & ~~~M_{M18} ~\cdots M_{M10} \cdot R\left(90^\circ\right)\cdot~M_{M9} \\
               & \cdot R\left(- (\mbox{Az} + 18.98^\circ) + 6.02^\circ\right) \cdot M_{M8} ~\cdots ~M_{M4} \\
               & \cdot R\left(90^\circ - \mbox{El}\right) \cdot M_{M3} \cdot R\left(\mbox{Pa}\right)
\end{aligned}
\end{equation}
As mentioned before, M$_1$ and M$_2$ are omitted as they can be neglected for the polarization analysis.

\subsection{Analyzing the input model}\label{sec:model_test}
\begin{figure}
	\centering
	\includegraphics[width=0.49\textwidth]{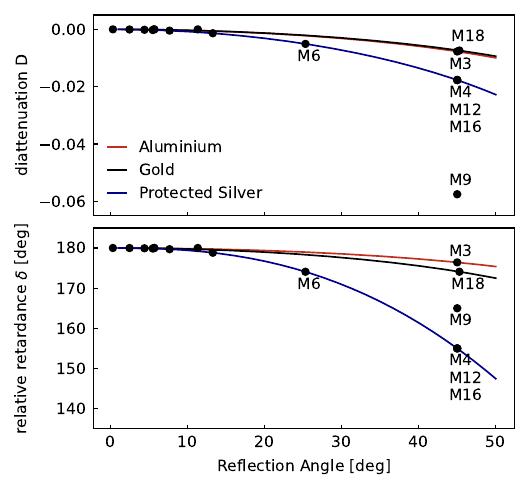}
	\caption{Diattenuation (top) and relative phase shift (bottom) of the individual VLTI mirrors in our model shown as a function of the reflection angle, for a wavelength of \SI{2250}{\nano\meter}. The worst offenders are labeled in both plots. The colored lines show the continuous values for the three used materials: Gold, Silver, and Aluminium. M9 does not lie on a line, as it is a dichroic mirror.}
	\label{fig:mirrors}
\end{figure}
\begin{figure*}
	\centering
	\includegraphics[width=0.95\textwidth]{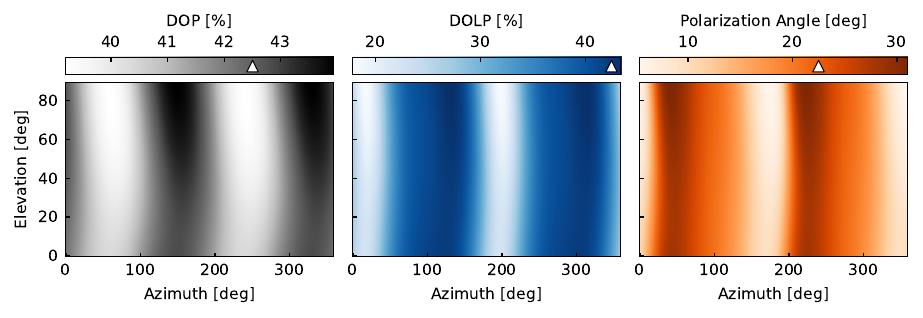}
	\caption{Properties of the output polarization for the input state $S = (1, 0.3, 0.3, 0)^T$ and all possible telescope positions. The plots show from left to right the degree of polarization (DOP), the degree of linear polarization (DOLP), and the polarization angle. The input state has a DOP and DOLP of \SI{42.5}{\percent} and a polarization angle of \SI{22.5}{\degree}. These values are shown as triangles in the color bars.}
	\label{fig:overview_model}
\end{figure*}
\begin{figure}
	\centering
	\includegraphics[width=0.36\textwidth]{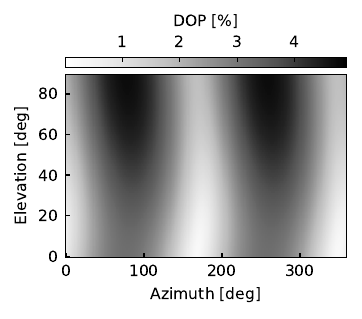}
	\caption{Expected DOP for unpolarized light at all telescope positions.}
	\label{fig:overview_model_IP}
\end{figure}
We will use calibration data to fit the input model in the next section, but we can use the theoretical model to understand some of the principal properties of the VLTI mirror train and how it will affect polarization. In \autoref{fig:mirrors}, the diattenuation and the phase shift of all the mirrors are shown. In an ideal case, the diattenuation would be zero (no instrumental polarization), and the phase shift would be \SI{180}{\degree} (no crosstalk) for each mirror. One can immediately see that the mirrors with large reflection angles are the worst offenders. In terms of instrumental polarization, the culprit is the dichroic mirror M9. As it is not a metal mirror and the values are measured individually, it does not lie on the lines of the three used metals. In terms of phase shift, the worst mirrors are the silver mirrors. This is due to the protective coating, which introduces a significantly stronger phase shift than a non-coated silver mirror would experience (see \autoref{app:protected}). The dominant mirrors here are the mirrors M4, M12, and M16. M4 is at the Nasmyth platform of the telescope, and M12 and M16 are in the VLTI delay line. This introduces another effect, as there is the telescope rotation of the azimuth angle between them. This rotation worsens the effect of crosstalk, as we have mirrors with strong crosstalk terms with a field rotation in between them. This rotation adds a strong correlation with the azimuth position of the telescope to the polarization effects. The second rotation on the telescope is the elevation rotation, but the only mirror on a different side of the rotation than the others is M3. As shown in \autoref{fig:mirrors}, M3 has, despite its \SI{45}{\degree} reflectance angle, comparably good polarization properties. We, therefore, expect fewer effects from the change in elevation than from the change in azimuth. A further effect we can infer from \autoref{fig:mirrors} is that while the values for the retardation are significantly different from the ideal value of \SI{180}{\degree}, the values for the diattenuation are closer to zero. We, therefore, expect, in general, more crosstalk than instrumental polarization from this model.

To verify these conclusions, we model the response of an arbitrary input Stokes vector. We used the vector $S = (1, 0.3, 0.3, 0)^T$. This corresponds to an input state with a degree of polarization of \SI{42.5}{\percent} and an equal amount of linear polarization. The polarization angle is \SI{22.5}{\degree}. This is the maximum source polarization one expects in astronomical sources in the near-infrared, but the exact vector is randomly chosen as an example. To see how this input vector propagates through the VLTI, we calculate the VLTI Mueller matrix for all telescope states and show the result for the output states in \autoref{fig:overview_model}. In a second test, we repeated the calculation for unpolarized light. The polarization degree for this case is shown in \autoref{fig:overview_model_IP}. From these plots, we can verify a couple of conclusions:
\begin{itemize}
    \item While there is a strong correlation with the Azimuth angle, there is only a small dependence on the telescope's elevation.
    \item As the polarization quantities are only defined over a \SI{180}{\degree} range, there is a repetition after \SI{180}{\degree} of azimuth rotation.
    \item The DOP only varies by approximately \SI{\pm 2}{\percent} from the input state, showing that instrumental polarization is, in comparison to crosstalk, not the dominant effect for highly polarized targets.
    \item We see large variations in DOLP (\SI{\pm 12}{\percent}) and polarization angle (\SI{\pm 13}{\degree}). This shows the large amounts of crosstalk in the VLTI path, shifting power between the Stokes parameters Q, U, and V.
    \item While the crosstalk dominates for highly polarized targets, we still expect some instrumental polarization. This is, depending on the telescope position, in the range of 0-\SI{4}{\percent}.
\end{itemize}
From this analysis, we conclude that our model produces output states which behave as expected and move on to calibrating it with test data.

\section{Calibration measurement}\label{sec:calibration}
To verify and calibrate our model, we took calibration data at the VLTI. As a light source, we used a high-power Thulium Laser from IPG photonics with a laser wavelength of \SI{1908}{\nano\meter}. As the instrumental polarization is expected to change with the wavelength,  we used a laser at \SI{1908}{\nano\meter} to be as close as possible to the science wavelength of GRAVITY (between 2000 and \SI{2500}{\nano\meter}). The differences over \SI{500}{\nano\meter} should be minor, as the refractive indices do not change significantly. In \autoref{app:model}, we analyze what differences our model predicts for this change in wavelength and confirm that the differences are minor compared to the uncertainties. Later, we can use calibration observations on sky to verify that it is not a limiting factor. The polarization measurements were done with a PAX polarimeter from Thorlabs. This is a rotating-waveplate-based polarimeter that Thorlabs customized to work at NIR wavelengths. To have the full light path of the VLTI, we launched the laser in the VLTI lab from the reference plates just in front of the GRAVITY feeding optics. With a linear polarizer, a half-wave, and a quarter-wave plate, we could modify the polarization of the laser and set it to arbitrary input states. The measurement head of the polarimeter was mounted onto a spider arm of one UT. This allowed us to measure the full light path at different telescope positions. For a sketch of the experimental setup, see \autoref{fig:sketch_exp}.

To get the complete polarization information, the goal was to measure the Mueller matrix of the light path at different telescope positions. As a Mueller matrix has 16 free parameters, we needed at least four input states to determine the entire matrix. As shown in \cite{Layden2012, Sabatke2000, Reddy2014}, it is best to equally space the input states over the possible parameter space, which minimizes the error from the matrix inversion. 

To keep our input states as simple as possible, we used six different input states with equal distribution over the Poincar\'{e} sphere. The input states were four linearly polarized states with a distance of 45 degrees from each other and two fully circularly polarized states (left and right). The linear states were intentionally chosen not to coincide with the geometric axis of the light path (i.e., not \SI{0}{\degree}, \SI{45}{\degree}, ...). The six input states we used are the following:
\begin{itemize}
\item 100 \% linear polarized at \SI{75}{\degree} 
\item 100 \% linear polarized at \SI{30}{\degree} 
\item 100 \% linear polarized at \SI{-15}{\degree} 
\item 100 \% linear polarized at \SI{-60}{\degree} 
\item 97 \% circular polarized, left-handed
\item 97 \% circular polarized, right-handed
\end{itemize}
The two circular states are only 97\% circular polarized, as it was difficult to get a state with zero ellipticity in our test setup. For better reproducibility, we settled on those states. The measurement was done for all four UTs with an average of 12 telescope positions in altitude and azimuth. To test how well we can measure the polarization and how reproducible the input states are, we did separate test measurements. In these measurements, we placed the polarimeter in the VLTI lab directly behind the optics to modify the polarization of the laser. By repeatedly going through all different input states in the same manner as for the telescope measurements, we could estimate the reproducibility of these states. We conclude that the uncertainty on the polarization angle is on the order of \SI{0.5}{\degree} and \SI{0.2}{\percent} for the degree of polarization. Those uncertainties are added to the error derived from the temporal scatter of each measurement.

\subsection{First results}
\begin{figure*}
	\centering
	\includegraphics[width=0.75\textwidth]{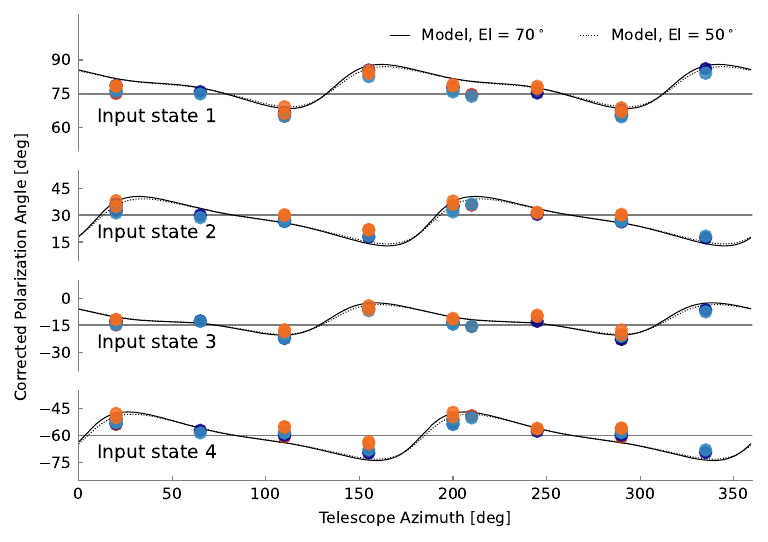}
	\caption{Measured polarization angles for the different telescope positions, only corrected by the geometric rotation of the field. The four panels show the data for each linear input state, with the different colors showing the data from the four UTs. The data from 50 and \SI{70}{\degree} elevation are shown together. The grey horizontal line shows the input values for the polarization angles, the black solid line the value expected from the model for \SI{70}{\degree} elevation, and the black dotted line the model for \SI{50}{\degree} elevation.}
 	\label{fig:pol_ang}
\end{figure*}
\begin{figure*}
	\centering
    \includegraphics[width=0.9\textwidth]{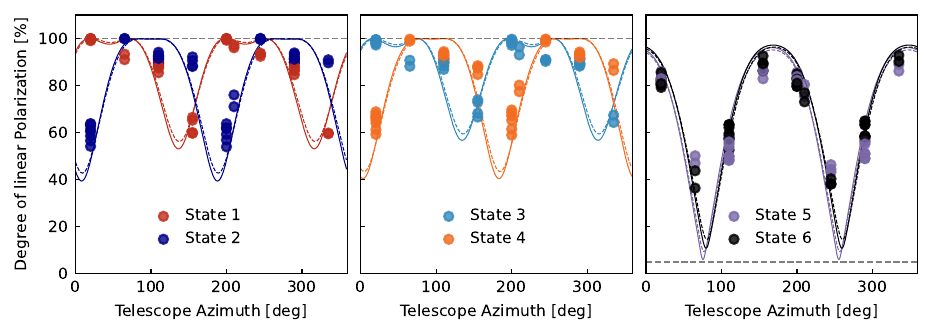}
    \caption{Measured degree of linear polarization for the different input states. Left and middle: linear input states where the input degree of linear polarization is 100\%. Right: circular states, where the input is fully circularly polarized, so 0\% linear polarization. The data points are shown in different colors for the different states. The data from 50 and \SI{70}{\degree} elevation are shown together. The lines in the same color show the model prediction for each state (solid line for \SI{70}{\degree} elevation and dotted for \SI{50}{\degree} elevation). The grey dashed line shows the input values in all plots.}
	\label{fig:dolp}
\end{figure*}
The first test we did was to check if the degree of polarization is maintained or if we have significant depolarization in the light path of the VLTI. Overall, we measure a degree of polarization of \SI{98.1(4)}{\percent} and, therefore, a polarization loss of around \SI{2}{\percent}. Such a small amount of depolarization is expected from scattering on dust in the optical train and again indicates that there is not a large amount of instrumental polarization.

More interesting is the effect of birefringence on the polarization angle and the degree of linear polarization. For the polarization angle, we look at the four linear input states, \SI{100}{\percent} linearly polarized at 75\si{\degree}, -60\si{\degree}, -15\si{\degree}, and 30\si{\degree}. This measurement must be corrected for the field rotation, according to \autoref{equ:fieldrot}, with an additional correction for the fact that the polarimeter measurement head was mounted on the telescope spider, which is at an angle of \SI{5.5}{\degree} from the central axis. 

The measured polarization angles as a function of azimuth position are shown in \autoref{fig:pol_ang}. These data were taken at an elevation of 50\si{\degree} and 70\si{\degree}. As the change in elevation is not the dominant factor, the data is shown in the plot together. This is only true for these figures. Later, each telescope state will be fitted with its correct telescope elevation. While the measured values lie around the input values, there is some modulation of around \SI{15}{\degree}. This is the crosstalk introduced by the mirror train, which clearly depends on the telescope's position. We see the same effects and the same order of magnitude here, as we have seen in \autoref{sec:model_test} and \autoref{fig:overview_model}. With the data, we also show the prediction of the polarization angle by the polarization model from \autoref{sec:model}. The model is shown for an elevation of \SI{70}{\degree }, as well as \SI{50}{\degree}, illustrating again that the change in azimuth introduces a more dominant effect than the elevation change. We also see clearly that there is a \SI{180}{\degree} ambiguity with the azimuth angle, which is, as discussed earlier, expected because the polarization properties are only defined in a range of \SI{180}{\degree}.

The figure generally illustrates that the data roughly follows the model, and the amount of crosstalk we measure is nicely predicted by the model. The data from all telescopes are shown in one plot to illustrate that the telescopes behave very similarly. The differences between the UTs will be discussed in \autoref{sec:differential}.

The third effect we can investigate is how much cross-talk there is between the linear and circular polarization states, i.e., how elliptic the input states become. The result is shown in terms of the degree of linear polarization in \autoref{fig:dolp}. Here, we see that the linear states, which should have \SI{100}{\percent} linear polarization, have much lower values, going down to below \SI{60}{\percent}, again depending on the telescope position. The inverse effect is clearly shown for the circular states, which reach very high values in the degree of linear polarization with a maximum of \SI{90}{\percent}.

From the calibration data, one can conclude that the UTs behave very similarly and do not show significant depolarization. However, we clearly see polarization effects, which would modify a polarization angle measurement by up to \SI{15}{\degree}. There is also substantial crosstalk between linear and circular states, which could decrease the measured linear polarization degree by up to \SI{40}{\percent}. Both effects are dependent on the telescope's position. At this point, we have not done any fitting yet, but we can already say that the model predicts the data very well.

\subsection{Fitting the calibration model}
\begin{figure*}[t]
	\centering
    \includegraphics[width=0.98\textwidth]{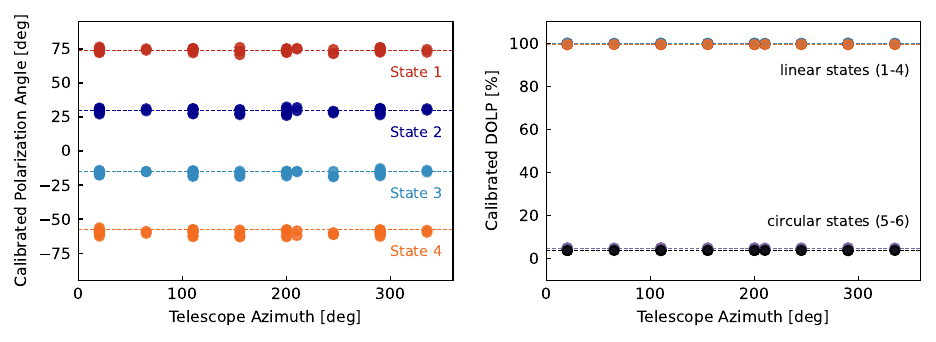}
	\caption{Test data calibrated with the fitted polarization model. The data are the same as in \autoref{fig:pol_ang} and \autoref{fig:dolp} but now calibrated. In both plots, the data are shown as dots, and the input states as dashed lines. The different colors show the different input states. The left plot shows the polarization angle and the right plot shows the degree of linear polarization. Since all states are fully polarized, the recovered degree of circular polarization is (100 - DOLP).}
	\label{fig:fit}
\end{figure*}
To reach a full calibration model, we improve the purely analytic VLTI model by fitting it to the obtained calibration data. The model includes 18 mirrors with two input values for the refractive indices and several rotations in the train. This has proven to be almost impossible to fit to our sparse data. To overcome this, we group all mirrors which have no rotation between them. These are the following:
\begin{itemize}
	\item M3
	\item M4 to M8
	\item M9
	\item VLTI lab and delay lines (M10 to M18)
\end{itemize}
The groups of mirrors are also indicated in \autoref{fig:sketch_exp} by different colors in the mirror notation. The groups rotate then with a change of elevation between M3 and M4, a change in azimuth between M8 and M9, and a constant field rotation after M9:
\begin{equation}
\begin{aligned}
    M_{VLTI} = & ~~~M_{Lab} \cdot R\left(90^\circ\right)\cdot~M_{M9} \\
               & \cdot R\left(- (\mbox{Az} + 18.98^\circ) + 6.02^\circ\right) \cdot M_{M4-8} \\
               & \cdot R\left(90^\circ - \mbox{El}\right) \cdot M_{M3} \cdot R\left(\mbox{Pa}\right)
\end{aligned}.
\label{equ:modelall}
\end{equation}
The advantage of this approach is that the form of the Mueller matrix for a group of reflections stays the same as for a single reflection (see \autoref{equ:mueller}). As the values in this matrix do not correspond to the values from a single Fresnel equation anymore, we can modify the Mueller matrix to the matrix in \autoref{equ:mueller_mod}. This leaves us with two quantities for each mirror group, the diattenuation and the phase shift. The values for M9 were measured by \cite{Lazareff2014}, which leaves us with only six values to fit. Furthermore, the fitted matrices do not include rotations, which makes it possible to apply the model for each telescope position and in both propagation directions.

\begin{table}
\caption{Fitted values for D and $\delta$.}        
\label{tab:fit}
\centering                          
\begin{tabular}{r c c}       
\hline\hline         
\rule{0pt}{2ex} \rule{0pt}{2ex} & D [$10^{-3}$] & $\delta$ [$^\circ$] \\
\hline 
\rule{0pt}{2ex} Values from fit: & \\
\rule{0pt}{2ex} M3 \rule{0pt}{2ex}  & -6.14 $\pm$ 4.7  &  170.0 $\pm$ 0.9\\
\rule{0pt}{2ex} M4-M8 \rule{0pt}{2ex}  & 2.58 $\pm$ 2.7 &  144.4 $\pm$ 0.3\\
\rule{0pt}{2ex} M10-M18 \rule{0pt}{2ex}  & 87.35 $\pm$ 1.9 &  142.3 $\pm$ 1.2\\
\hline
\rule{0pt}{2ex} Values from model: & \\
\rule{0pt}{2ex} M3 \rule{0pt}{2ex}  & -7.7  &  176.4 \\
\rule{0pt}{2ex} M4-M8 \rule{0pt}{2ex}  & 24.7 &  147.5 \\
\rule{0pt}{2ex} M9 (dichroic) \rule{0pt}{2ex}  & 57.47 &  165.0 \\
\rule{0pt}{2ex} M10-M18 \rule{0pt}{2ex}  & 43.0 &  123.9 \\
\hline
\end{tabular}
\tablefoot{The given uncertainty is derived by bootstrapping the full data set, fitting the data of the four telescopes individually, and calculating the standard deviation over the four values. For the final uncertainty, the larger value of the two methods is chosen. The values for M9 are not fitted and, therefore, have no uncertainty. For comparison, the lower part of the table shows the values expected from the model without a fit.}
\end{table}

With the fitted values, we, therefore, have a polarization model that calculates a Mueller matrix for the whole VLTI light path and depends on the telescope position. For an ideal mirror, we assume $D=0$ and $\delta=180^\circ$. The values derived from the fit are listed in \autoref{tab:fit}. The uncertainties are derived from the scatter of fitting each telescope individually and from bootstrapping the dataset. The results for the instrumental polarization have a significant error bar relative to the values, but the values are all on the order of $10^{-2}-10^{-3}$, which again shows that there is only a little instrumental polarization. The values for the phase shift differ more strongly from $\delta=180^\circ$, with the mirrors in the lab and the delay line (M10-M18) and the mirrors in the telescope (M4-M8) contributing equally. This again confirms the findings from the model in \autoref{sec:model_test}. The comparable values for $\delta$ for both parts are expected, as we showed in \autoref{fig:coating}, M4, M12, M16, and less strongly M6 should have the most significant impact on the retardance. The retardance of several mirrors adds up, and given that the worst offenders are situated in both parts of the train, one would not expect one part to be significantly better than the other. From our model, we would have expected that the lab and delay line part would be worse as it contains M12 and M16, but the fitting results show a comparable phase shift for the group M4-M8. This group contains exclusively protected silver coatings, and as mentioned earlier, they might have different protective coatings and are more challenging to model. We assume this is the case for the strong retardance of this group.

With the fitted values, the calibration model is a simple function of telescope position. We obtain a Mueller matrix for each telescope position, which describes the instrumental polarization of the VLTI, by executing \autoref{equ:modelall}. The sky polarization can be calculated from the measured Stokes vector and the Mueller matrix of the VLTI by applying \autoref{equ:smueller}.

With the fitted calibration model, we calibrate the test data set, shown in \autoref{fig:fit}. We obtained the Mueller matrix of each telescope position from our calibration model and multiplied it to the data. For the calibrated data, the polarization angle of the input states is recovered well, and the degree of linear polarization is \SI{100}{\percent} for the linear input states and very low for the circular states, which matches the input states. If one compares the calibrated data with the original in \autoref{fig:pol_ang} and \ref{fig:dolp}, this is a very clear improvement. The calibrated polarization states match the input states with an average error of \SI{0.5}{deg} in the polarization angle and \SI{0.4}{\percent} in the degree of linear polarization. This excellent agreement of the fitted model also validates the simplifications made to the mirror train in the model (see \autoref{sec:modeling}).

\section{Instrumental polarization of GRAVITY}\label{sec:gravity}
So far, the results have been independent of the interferometric instrument and generally valid for VLTI observations in the near-infrared. However, to calibrate polarized observations, the instrument has to be taken into account as well. Here, we discuss the instrumental polarization of the GRAVITY beam combiner. For a complete overview of GRAVITY, see \cite{GRAVITY2017}.

\subsection{GRAVITY light path}\label{sec:gravity_path}
\begin{figure}
	\centering
    \includegraphics[width=0.45\textwidth]{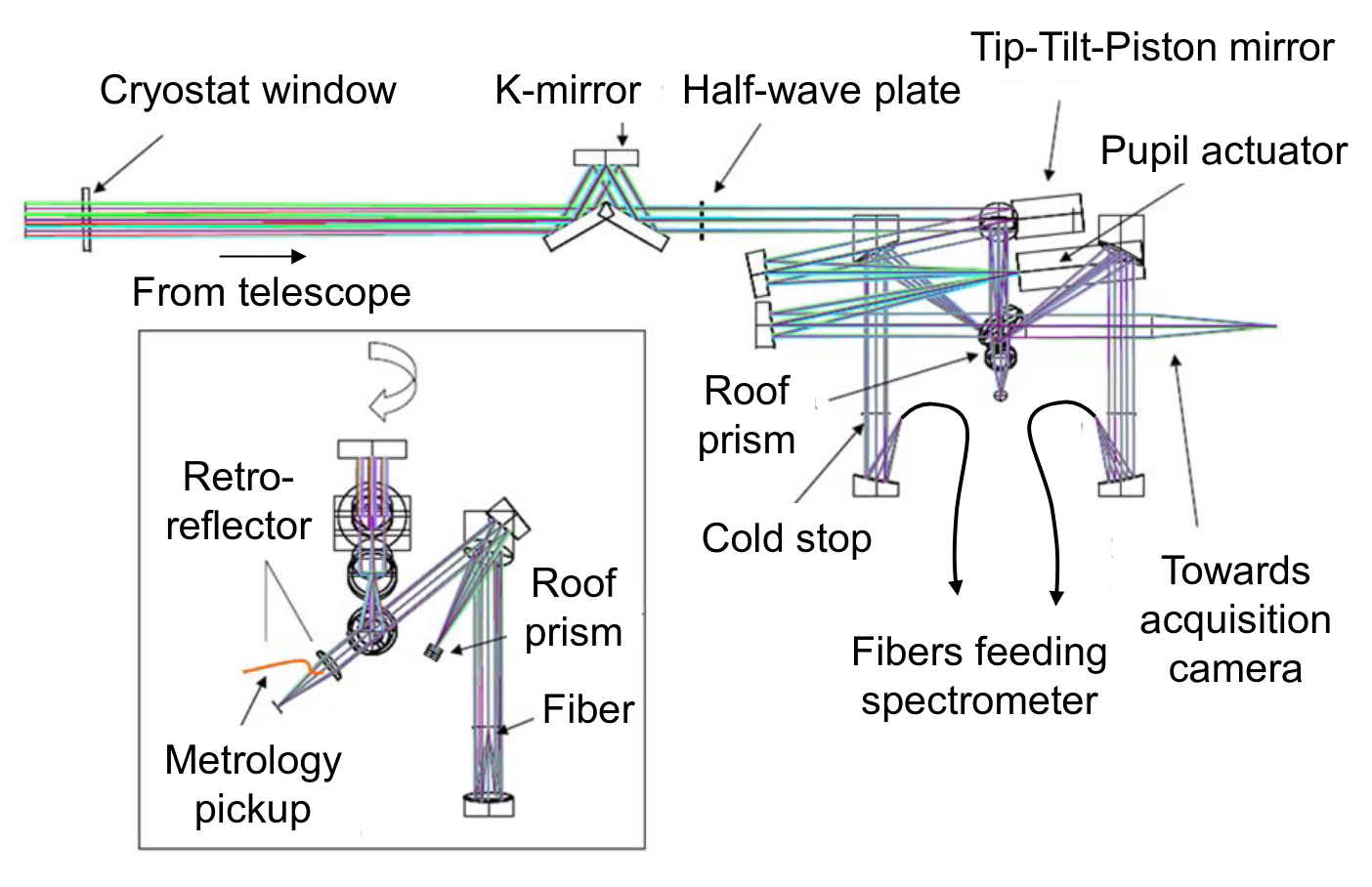}
	\caption{Schematic side and front view of the optical design of the GRAVITY fiber coupler. From \cite{GRAVITY2017}.}
	\label{fig:coupler}
\end{figure}
In GRAVITY, the light first passes the fiber coupler \citep{Pfuhl2014}. Part of the fiber coupler is a K-mirror to de-rotate the field and a half-wave plate (HWP). The K-mirror, as well as the HWP, rotate in a fixed way during the observation: The K-mirror is used as a derotator and moves according to the field rotation described in \autoref{equ:fieldrot}. With the derotation of the field, it also derotates the sky polarization. Normally we would, therefore, not need to derotate the polarization signal. However, the HWP in GRAVITY, which follows the K-Mirror in the light path, rotates opposite the K-Mirror and reintroduces the field rotation in the polarization. The reason for this is that it allows for the metrology laser, which backpropagates through the lightpath, to have a stable polarization. This laser follows the full light path of GRAVITY and the VLTI before it is used to measure differential optical path differences above the primary mirror. This path difference is measured by using the interference between the light from the science and the fringe tracker beam. This measurement allows for phase references astrometry with GRAVITY but is extremely sensitive. To get the best possible contrast in the interference pattern, the polarization of the metrology beam is kept stable in the VLTI, and therefore, the polarization of the starlight in GRAVITY rotates with the field. The rotation correction described in \autoref{sec:fieldrot} and \autoref{equ:fieldrot} must still be applied to a polarization measurement. 
\begin{figure}
	\centering
    \includegraphics[width=0.45\textwidth]{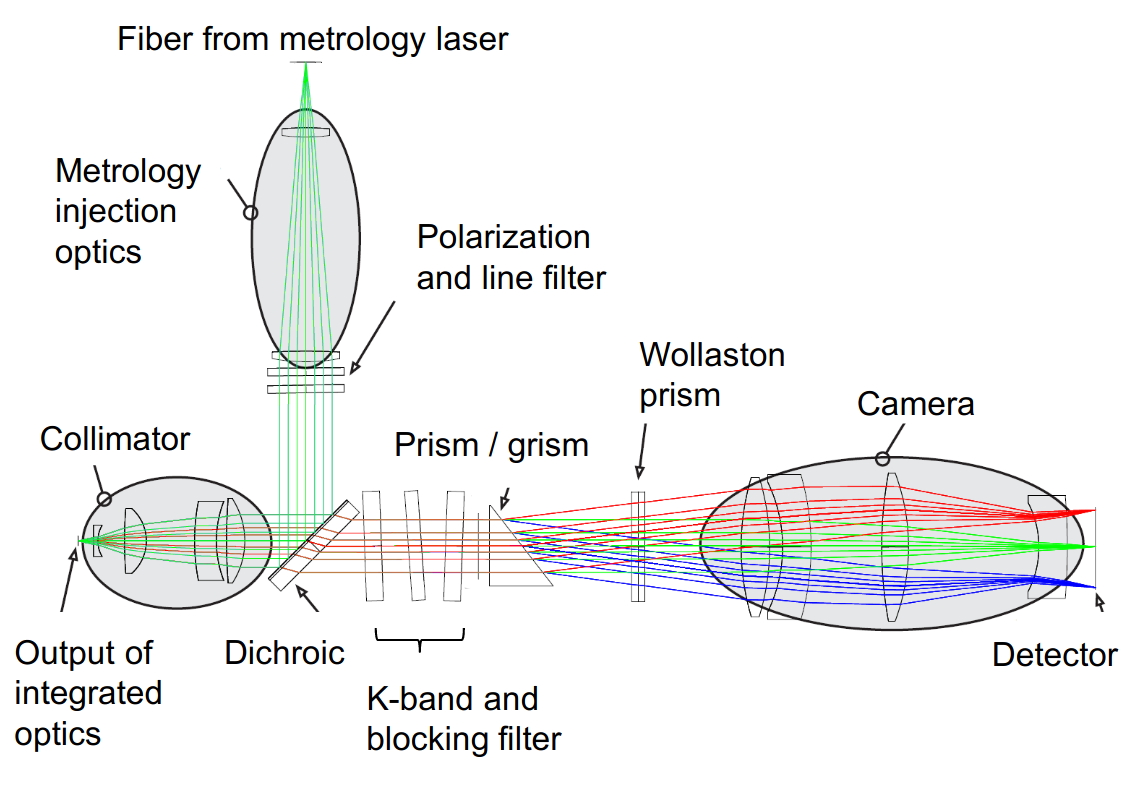}
	\caption{Optical design of the spectrometer. Most important for the polarimetric mode is the Wollaston prism, which can be moved in or out of the light path. From \cite{GRAVITY2017}.}
	\label{fig:spectrometer}
\end{figure}
After the K-Mirror and the HWP follow the tip-tilt, piston, and pupil control before the light is split into science and fringe-tracker and fed into optical fibers (\autoref{fig:coupler}. In the fibers, it passes the fiber control unit. Part of this unit are Fibered Polarization Rotators, which rotate the polarization direction in the fiber and are used to match the polarization of all baselines \citep{GRAVITY2017}. From the optical fibers, the light is fed into the integrated optics system \citep{Jocou2014, Perraut2018} and finally passed into the spectrometers \citep{Straubmeier2014}. In the spectrometers, there are Wollaston prisms, which can be put into the light path to allow for a polarimetric measurement (see \autoref{fig:spectrometer}) and split up the light into two polarizations (P1 and P2), with a \SI{90}{\degree} polarization angle between them. The polarization P1 is horizontally polarized in the VLTI lab frame or aligned with V (in the general Paranal coordinate system (V, W), where V is horizontal and W vertical in the lab \citep{Gitton2009}).  GRAVITY also includes a calibration unit \citep{Blind2014}, which can be used to test and calibrate the instrument. For this, it creates artificial stars in all beams. The calibration unit also includes the option to use a linear polarization filter to fully polarize the artificial light sources.

There are some field rotations in the light path of GRAVITY. However, the field rotations do not change over time. The GRAVITY fiber-coupler is aligned to ensure that a horizontal polarization on the calibration unit corresponds to one of the polarization directions on the detector. For this alignment, the linear polarizer in the calibration unit is used. With the linearly polarized light from the calibration unit, the Fibered Polarization Rotators in the fiber control unit of GRAVITY are optimized to get a fully illuminated P2 spectrum on the detector, and no light on the P1, showing that the polarization vector is aligned with the vertical axis on the detector. The field rotations inside GRAVITY are, therefore, compensated by this alignment and don't have to be taken into account in this model.
The Fibered Polarization Rotators are made out of standard, not polarization-maintaining fibers, that can be rotated to adjust the polarization angle. The effect of this rotation is only to rotate the polarization. The fibers themselves are weakly birefringent, and no effect of the rotators on birefringence has been measured \citep[see][]{Perrin2023}. The polarization effects of this system are, therefore, included in the measurement of GRAVITY and do not change with time.

\subsection{Measurements of polarization effects}
\begin{figure*}
	\centering
    \includegraphics[width=0.8\textwidth]{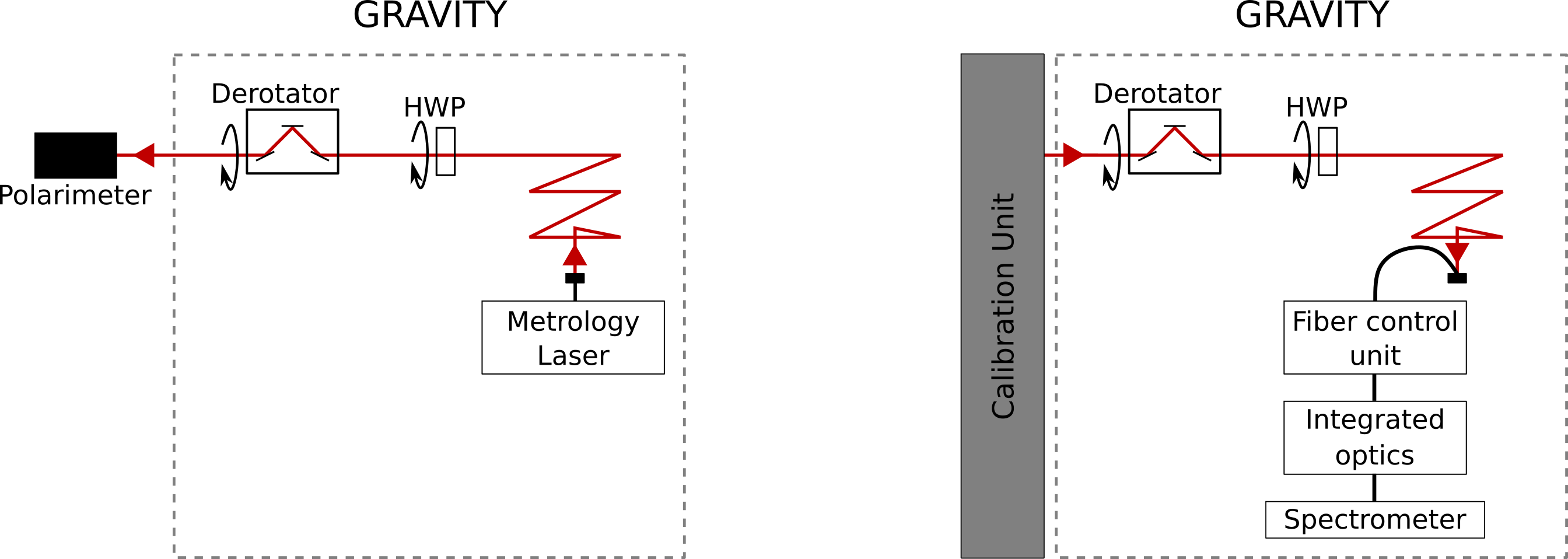}
	\caption{Sketch of the two polarimetric measurements for GRAVITY. Left: In the first measurement, the light from the metrology laser is used and measured with a polarimeter. Right: The light from the calibration unit is used and recorded on the spectrometer. In both cases, the light direction is indicated by arrows. The light path is the same as shown in \autoref{fig:coupler}, with the reflection for the Tip-Tilt and Piston control just shown as a change in direction in the light path.}
	\label{fig:grav_measurement}
\end{figure*}
To measure the polarization effects of GRAVITY, we performed two individual experiments. The first one was done with the same polarimeter as the VLTI measurements. We put the polarimeter in front of one of the beams and used the metrology laser as a light source. The second experiment uses the light of the calibration unit, which can be used with or without a linear polarization filter, producing fully polarized or unpolarized light. For the experiment with the calibration unit, we measure the signal on the detector, which can only measure one Stokes parameter at a time. To get the full linear polarization, we need, therefore, to rotate the HWP between two exposures. This assumes that the HWP behaves as it should. This is where we use the polarimeter experiment to confirm that this is the case. A sketch of the two experiments is shown in \autoref{fig:grav_measurement}, and we will discuss them in detail in the following.

\subsubsection{Polarimeter measurement}
\begin{figure}
	\centering
    \includegraphics[width=0.45\textwidth]{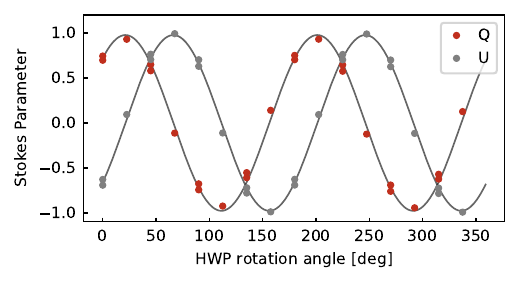}
	\caption{Measurement of the polarization in GRAVITY with a rotating HWP. The points show the data points for Stokes Q (red) and Stokes U (grey). The black lines show the corresponding values for an ideal optical element.}
	\label{fig:grav_pax}
\end{figure}
In the first measurement, we used the metrology laser as a light source. The metrology laser in GRAVITY is split into three parts: two low-power parts (with less than 1 \% of the laser power), which follow the full light path, and one high-power part fed into the light path only after the fiber optics. We used this last part, the so-called carrier beam, and measured it with the PAX polarimeter outside GRAVITY. The carrier beam follows some mirror optics and then passes the HWP and the derotator before leaving GRAVITY (as indicated in the left part of \autoref{fig:grav_measurement}). We used the polarimeter of the outcoming light for each beam and rotated the half-wave plate. The measured Stokes Q and U values are shown in \autoref{fig:grav_pax}. 

The carrier beam we use in this measurement is fully polarized, with a fixed polarization state. However, the polarization direction was only coarsely aligned during the integration of GRAVITY, so we only have a vague idea about the polarization angle. Furthermore, the carrier's light is fed into the light path at the fiber coupler, and we, therefore, do not have a measurement of some parts of the instrument, mainly the fibered and integrated optics. We cannot use this measurement to characterize GRAVITY fully, but we can use it to characterize the HWP. The measured Stokes parameters are shown in  \autoref{fig:grav_pax} as a function of HWP rotation angle. As solid lines, the response of a perfect HWP is shown. The measured Stokes parameters show a very good agreement with the theoretical expectation, with an average discrepancy of 0.005. Given that small value in comparison to the other uncertainties in the polarization calibration, we decided to treat the HWP as an ideal component and will describe it with the following Mueller matrix:
\begin{equation}
\label{equ:hwp}
M = \left( 
\begin{array}{cccc}
1 & 0 & 0 & 0 \\
0 & 1 & 0 & 0 \\
0 & 0 & -1 & 0 \\
0 & 0 & 0 & -1 
\end{array}
\right).
\end{equation}
For all other elements of GRAVITY, we refer to the results from the second measurement.

\subsubsection{Calibration unit measurement}
\begin{figure*}
	\centering
    \includegraphics[width=0.8\textwidth]{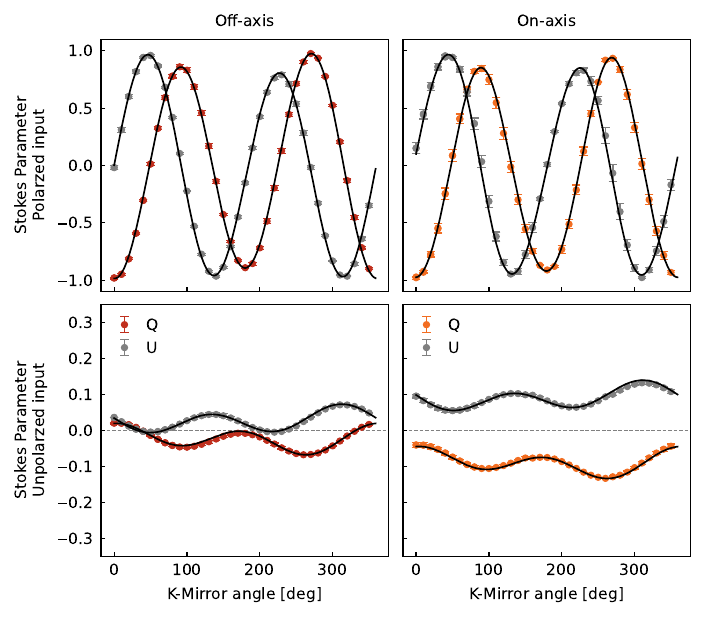}
	\caption{Measured polarization with GRAVITY in the different observing modes. The left column shows the measurement in the off-axis mode, and the right column in the on-axis mode. In the top row, the linear polarization filter is used for the input light source; in the bottom, it is not. In all plots, the Stokes Q data points are shown in red/orange and the Stokes U in grey. The data is the average over all four GRAVITY beams. The results from the fitted model are shown in black lines.}
	\label{fig:grav_full}
\end{figure*}
In the second experiment, we used the light from the calibration unit of GRAVITY, which is a Quartz Tungsten Halogen lamp with the possibility to add a linear polarization filter. Using this lamp as a light source, with the polarization filter, we get a constant and linearly polarized input source. We then rotated the K-Mirrors of each beam from its initial position to \SI{360}{\degree}. At each location of the K-Mirror, we took detector frames with the HWP at 0 and \SI{22.5}{\deg}. The mechanical rotation of \SI{22.5}{\deg} corresponds to a rotation of the polarization angle by \SI{45}{\deg}.  At each position, we extracted the measured flux from the detector, split up into P1 and P2 by the Wollaston prism. Following \autoref{equ:stokes_measure} allows for measuring the linear stokes parameter. With this experiment, we measure the polarization on the GRAVITY science detector for a linear input polarization with a rotating polarization angle, shown in the top panels of \autoref{fig:grav_full}. As there is no quarter-wave plate in GRAVITY, there is no possibility to measure stokes V, but only the linear polarization parameters Q and U. As this leaves us with an incomplete measurement for the Mueller matrix of GRAVITY, we repeated the full measurement, but without the linear polarization filter of the calibration unit. The input light is then unpolarized and allows us to measure the instrumental polarization of GRAVITY. The data are shown in the bottom panels of \autoref{fig:grav_full}. 

GRAVITY has two different observing modes, which come with slightly different optical paths. The first, the off-axis mode, is the mode in which two different objects are observed as science (SC) and fringe-tracker (FT) objects. In this mode, the field is spatially split into two fields by a roof prism in GRAVITY. The second mode is the on-axis mode, in which the same object is used for SC and FT. In this case, the light is split by a beam splitter in GRAVITY. Apart from the roof and the beam splitter, the light path is identical for both modes in GRAVITY. However, as these two elements are distinctly different, one has to assume that this affects polarization. Therefore, we did the full calibration unit test twice, once for the on- and for the off-axis mode. This leaves us with four measurements in total: the polarized and unpolarized input in both on- and off-axis mode. All four measurements are shown in \autoref{fig:grav_full}. We repeated these measurements for all four GRAVITY beams, and as the results were very similar, they are shown as averaged datapoints in \autoref{fig:grav_full}.

\paragraph{First look at calibration unit measurement}\mbox{}\\
Looking at the data with polarized input light (top row in \autoref{fig:grav_full}), one sees that GRAVITY in first order behaves as expected: The light is almost entirely linearly polarized, and the polarization direction rotates with the rotation of the K-Mirror. For the unpolarized input (bottom row in \autoref{fig:grav_full}), the measured polarization is much lower but not zero. Here, we see the main difference between the off-axis and the on-axis mode. In off-axis, the measured linear polarization for unpolarized input light is between 0 and \SI{8}{\percent}. In on-axis mode, we measure 8 to \SI{16}{\percent} for the same input light. We, therefore, have a much higher instrumental polarization in the on-axis mode, which can be explained by the fact that beam splitters often show non-ideal polarization properties.

\paragraph{Fit to the data}\mbox{}\\
To fit a model to the test data, we separately fit a Mueller matrix for the K-Mirror and GRAVITY. The HWP is assumed to be perfect, as motivated before:
\begin{equation}
\label{equ:fit_gravity}
\begin{aligned}
    M_{GRAVITY} = & M_{GR} \cdot R(\Theta_{HWP}) \cdot M_{HWP} \cdot R(-\Theta_{HWP}) \cdot \\
                  & R(\Theta_{KM}) \cdot M_{KM} \cdot R(-\Theta_{KM})
\end{aligned}
\end{equation}
Where $M_{GR}$ is GRAVITY without the HWP and the K-Mirror. As discussed before, the polarization properties of the on-axis and off-axis modes of GRAVITY are different. One, therefore, has to choose $M_{GR, on}$ or $M_{GR, off}$ here, depending on the observing mode. For simplicity, we will use $M_{GR}$ as a general name for it. Furthermore,  $R(\Theta_{HWP})$ and $R(\Theta_{KM})$ are the rotation matrices for the HWP and the K-Mirror. The K-Mirror consists of three individual mirrors, which are fixed in one mount that moves them together. We can describe them as one Fresnel reflection according to \autoref{equ:mueller_mod}. This leaves only two fitting parameters (diattenuation and retardation). The fit for the remaining optics of GRAVITY is more complicated. There are a large number of field rotations inside GRAVITY. They are all fixed and do not change with time, but this prohibits us from using a simple formula as for the K-Mirror. We could split GRAVITY into individual mirror groups, as we do for the VLTI, with a rotation relative to each other. However, this would not lead to a decrease in fitting parameters. Additionally, we do not need the freedom to fit the parameters individually, as the rotations do not change. So for GRAVITY, we just fit a full Mueller matrix (according to \autoref{equ:schem_mueller}) with only the first component being fixed to 1 as we look at normalized Stokes vectors. All other values can take values between -1 and 1. 

In summary, we are using the data from \autoref{fig:grav_full} and fit \autoref{equ:fit_gravity} to it by fitting each component of $M_{GR}$, as well as the diattenuation and retardation $M_{KM}$, while not fitting $M_{HWP}$, but using \autoref{equ:hwp} for it. As an initial guess, we take a perfect mirror for the K-Mirror, and we calculate the expected Mueller matrix for the rest according to the mirror positions and materials in GRAVITY. The fit is then done twice for the off as well as on-axis modes separately. 

The result of the fit is shown in \autoref{fig:grav_full}, and the resulting Mueller matrices can be found in \autoref{app:matrices}. The comparison to the data in \autoref{fig:grav_full} shows that the fit worked well, and the instrument is well described by the given matrices.

\section{Full calibration of VLTI \& GRAVITY}\label{sec:full_calib}
With the results of the GRAVITY measurement, we have all the information in hand to calculate the complete polarimetric response for an observation with GRAVITY and the VLTI. From the work presented in \autoref{sec:calibration} and especially \autoref{equ:modelall} we obtain the Mueller matrix of the VLTI, which depends on the elevation and azimuth of the telescope and is entirely defined by the parameters in \autoref{tab:fit}. Together with the Mueller matrix of the K-Mirror and the rest of GRAVITY, which are listed in \autoref{app:matrices}, the complete polarization response of GRAVITY and the VLTI is given by the following equation:
\begin{equation}
\begin{aligned}
    M_{ALL} = & M_{GR} \cdot R(\Theta_{HWP}) \cdot M_{HWP} \cdot R(-\Theta_{HWP}) \cdot \\
              & R(\Theta_{KM}) \cdot M_{KM} \cdot R(-\Theta_{KM}) \cdot M_{VLTI}(Az, El, pa),
\end{aligned}
\label{equ:fullmmatrix}
\end{equation}
where R is the usual rotation matrix (\autoref{equ:rotmat}) with the position of the HWP ($\Theta_{HWP}$) and K-Mirror ($\Theta_{KM}$). For $M_{GR}$, one has to choose the on- or off-axis one, matching the observing mode.

\section{Polarimetric measurements with GRAVITY}
\begin{table}
\caption{Necessary header keywords for the creation of the Mueller matrix.}
\centering
\begin{tabular}{r c}
\hline\hline     
\rule{0pt}{2ex} \rule{0pt}{2ex}
Value & Header Keyword \\ 
\hline
Azimuth & ESO ISS AZ \\[2pt]
Elevation & ESO ISS ALT \\[2pt]
Paralactic angle & ~~(ESO ISS PARANG START\\
& + ESO ISS PARANG END)/2 \\
K-Mirror position & $\frac{1}{4}\sum_{i=1}^4$ (ESO INS DROTi START\\
& \hspace{20pt} + ESO INS DROTi END)/2\\[2pt]
HWP position & $\frac{1}{4}\sum_{i=1}^4$ [(ESO INS DROT(i+4) START\\
& \hspace{24pt} + ESO INS DROT(i+4) END)/2]\\
\hline
\end{tabular}
\tablefoot{For the K-Mirror and the HWP angle, slight differences exist between the four beams, which are therefore averaged.}
\label{tab:header}
\end{table}

To observe with GRAVITY in a polarimetric mode, one has to put the Wollaston prism into the light path, which can be selected in the preparation of the observing block. This gives measurements in two polarizations, P1 and P2, as discussed in \autoref{sec:gravity}. From the interferometric signal in both polarizations, one can calculate the measured source intensity and the first linear Stokes parameter q:
\begin{equation}
    I = I_{P1} + I_{P2}, ~~~~ q = I_{P1} - I_{P2}
\label{equ:Q}
\end{equation}
To retrieve the second linear stokes parameter, u, one has to rotate the HWP \SI{22.5}{\degree}, which rotates the polarization axis by \SI{45}{\degree}:
\begin{equation}
    u = I^{22.5}_{P1} - I^{22.5}_{P2}
\label{equ:U}
\end{equation}
The rotation of the HWP can also be selected in the preparation of the observation and is added as an offset to the nominally calculated HWP position. This means that the HWP still follows its calculated position (see \autoref{sec:gravity_path}) but in the second case with an added offset of \SI{22.5}{degrees}.

With this q and u, one usually builds up a measured stokes vector and corrects it with the instrument matrix. However, this only works if the HWP is the last optical element in the beam. In the case of GRAVITY, the HWP sits at the entrance of GRAVITY, which means several optical elements after the HWP, which introduce birefringence. Therefore, what the detector sees is not the U state of the light but a Q with an altered instrument setup. We, therefore, assume we have measured Q twice, once with the HWP at \SI{0}{\degree} and one at \SI{22.5}{\degree}. In normalized Stokes parameters, the measurement is then given by:
\begin{equation}
Q_1 = \frac{I_{P1}^{0} - I_{P2}^{0}}{I_{P1}^{0} + I_{P2}^{0}}, ~~~~~ Q_2 = \frac{I_{P1}^{22.5} - I_{P2}^{22.5}}{I_{P1}^{22.5} + I_{P2}^{22.5}}
\label{equ:both_q}
\end{equation}

Usually, $Q_1$ and $Q_2$ are taken in two subsequent exposures. One then needs a separate Mueller matrix for both measurements. This matrix is given by \autoref{equ:fullmmatrix}, which then includes the VLTI and GRAVITY, with all field rotations and polarimetric effects. It depends on the Azimuth and Elevation of the telescope, on the paralactic angle, and the position of the HWP and K-Mirror. All this information is taken from the header of a normal GRAVITY fits file, with the keywords listed in \autoref{tab:header}. $M_{ALL}$ has to be calculated twice for the two exposures used for $Q_1$ and $Q_2$ in \autoref{equ:both_q}. Using the header values for the two subsequent exposures gives two Mueller matrices, $M_{ALL,1}$ and $M_{ALL,2}$ (The HWP offset in case two is already taken into account in the HWP values in \autoref{tab:header}).

With M describing the response of the system in $S_i = M_i \cdot S_{sky}$ and the components of M from \autoref{equ:mueller}, the measured $Q_i$ can be written as:
\begin{equation}
\begin{aligned}
    Q_i = & ~(I\rightarrow Q)_i \cdot I_{sky} + (Q\rightarrow Q)_i \cdot Q_{sky} \\
          & + (U\rightarrow Q)_i \cdot U_{sky} + (I\rightarrow V)_i \cdot V_{sky}
\end{aligned}
\end{equation}
Furthermore, we can define the polarization vector on sky as $S_{Sky} = (1, Q, U, 0)$. I is set to one as we are considering normalized Stokes parameters. We can set V to zero as sources in the near-infrared usually do not show significant circular polarization. This changes the given expression to:
\begin{equation}
    Q_i = (I\rightarrow Q)_i + (Q\rightarrow Q)_i \cdot Q_{sky} + (U\rightarrow Q)_i \cdot U_{sky}
\end{equation}
Separating the instrumental polarization and the crosstalk then allows the set up of a system of equations:
\begin{equation}
\begin{aligned}
\left[\begin{array}{c}
                     Q_1 - (I\rightarrow Q)_1\\
                     Q_2 - (I\rightarrow Q)_2\\
\end{array}\right]
& = 
\left[\begin{array}{c}
                     (Q\rightarrow Q)_1  ~~ (U\rightarrow Q)_1\\
                     (Q\rightarrow Q)_2  ~~ (U\rightarrow Q)_2\\
\end{array}\right]
\cdot
\left[\begin{array}{c}
                     Q_{sky}\\ 
                     U_{sky}\\
\end{array}\right]\\
Q_{vec} & = A \cdot \left[Q_{sky}, U_{sky}\right]^T
\end{aligned}
\label{equ:pol_leastsquare_in}
\end{equation}
Which can be solved using linear least squares:
\begin{equation}
\left[Q_{sky}, U_{sky}\right]^T = (A^TA)^{-1}A^TQ_{vec}.
\label{equ:pol_leastsquare}
\end{equation}
This gives a measurement of Q and U on sky for every two exposures. In case the source polarization is assumed to be constant over time, \autoref{equ:pol_leastsquare_in} can easily be extended with more than two Q measurements \citep[see][]{vanHolstein2020}. Another solution to this approach is to forward model the polarimetric property of the source \citep{Gravity2020pol} to recover the complete polarization information. In this case, $M_{ALL}$ can just be taken from \autoref{equ:fullmmatrix}.

\section{Application to data}\label{sec:data}
\begin{table}
\caption{Polarization of degree (P) and angle ($\Theta$) for IRS~16C}             
\label{tab:irspol}
\centering                          
\begin{tabular}{r c c}       
\hline\hline         
\rule{0pt}{2ex} \rule{0pt}{2ex} & P [\%] & $\Theta$ [$^\circ$] \\
\hline 
\rule{0pt}{2ex} \cite{Ott1999} \rule{0pt}{2ex}  & 4.0 $\pm$ 1.6 & 35 $\pm$ 19 \\
\cite{Witzel2011} & 4.6 & 17.8 \\
\cite{Buchholz2013} & 4.3 $\pm$ 0.6 & 25 $\pm$ 5 \\
 & & \\
\textbf{This work:}  & 4.3 $\pm$ 0.4 & 19.9 $\pm$ 4.6 \\
\hline       
\end{tabular}
\end{table}
\begin{figure*}
	\centering
    \includegraphics[width=0.75\textwidth]{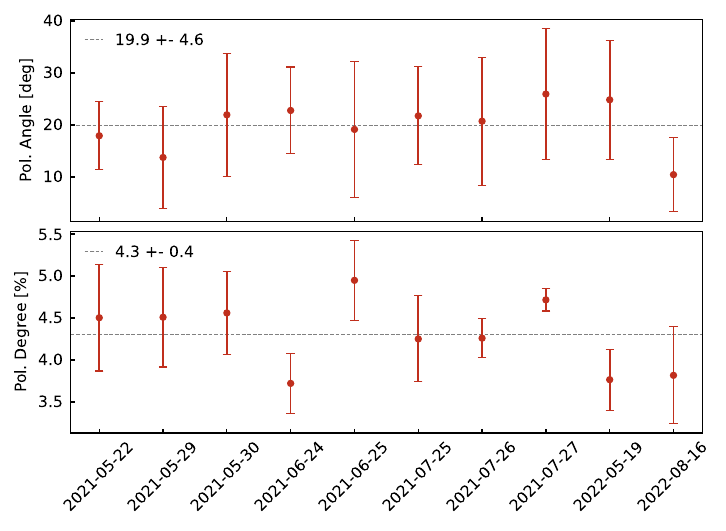}
	\caption{Polarization measurements of IRS~16C between 2019 and 2022. The top panel shows the polarization angle for each night, and the bottom panel shows the polarization degree. The average over all nights is shown as a horizontal line}
	\label{fig:alldata}
\end{figure*}
To test the calibration model, we use data taken with GRAVITY of the Galactic Center. For a description of the observation of the Galactic Center black hole and its surrounding stars, see, for example, \cite{gravity2018redshift, GRAVITY2018, gravity2020flux}. As shown in those papers, the science target Sgr~A* is not a good test target as it has a variable polarization state. However, the observations in the Galactic Center are done in the off-axis mode. In this mode, a close-by single star is used as a phase-reference source \citep[for more information see][]{GRAVITY2017}. In the case of the Galactic Center, the phase reference source is the star IRS~16C, which has a brightness of m$_K$ = 9.55 and is well known to be slightly polarized. It was, for example, observed in \cite{Ott1999, Witzel2011, Buchholz2013}, and the studies found a consistent polarization of \SI{4}{\percent} (see \autoref{tab:irspol}) due to foreground polarization from dust.

In the off-axis mode, GRAVITY acts as two individual interferometers, which are referenced to each other via a metrology system. While the goal of the measurement was to measure the polarization of Sgr~A*, we equally get an interferometric observation of the phase reference target. As the rotation of the HWP affects this target in exactly the same way as it does the science target, we apply the full analysis to the measured fluxes of the fringe-tracking star to measure its polarization. As this is a single star, one can simply take the coherent flux calculated by the GRAVITY data reduction software (DRS). We use the flux per polarization channel, take the average over the spectrum and all telescopes, and calculate $Q_i$ for each exposure following \autoref{equ:both_q}. We then get $Q_{sky}$ and $U_{sky}$ by using \autoref{equ:pol_leastsquare_in} and \autoref{equ:pol_leastsquare}.

As IRS~16C is the usual fringe-tracking target for Galactic Center observations, there are several data sets with polarization observations of the star. \autoref{fig:alldata} shows the polarization angle and degree for different nights between 2019 and 2022. The polarization state in each night is well measured with an average scatter of the polarization angle of \SI{5}{\degree} and \SI{0.4}{\percent} for the polarization degree per night. This shows that the polarization calibration works well for different telescope orientations and that we can measure reliable polarization states with GRAVITY. With the complete analysis, we measure a polarization degree of (4.3 $\pm$ 0.4)\si{\percent} at an angle of  (19.9 $\pm$ 4.6)\si{\degree}. These values perfectly agree with the other values from \autoref{tab:irspol}. 

The fact that we can measure the expected polarization of the fringe-tracking star verifies our model and the polarimetric capabilities of GRAVITY. It also confirms the results from \cite{Gravity2020pol}, where we already used the presented calibration model and could show clear and varying signals in the polarization of Sgr~A*.

The way to extract the flux in the different polarization states from GRAVITY data depends on the observed target. While for single targets, such as IRS~16C, the DRS outputs can be used, a more complicated source structure might involve a different analysis. For the case of Sgr~A*, for example, we apply a multi-source fit, which fits a model of the central black hole and several surrounding stars to the data, including the flux ratio between the stars and the black hole. By doing this for both polarizations individually, we can measure the flux ratio with respect to nearby stars in each polarization \citep[for more details see][]{gravity2020flux}. The flux measurement in the different polarization states will, therefore, depend on the science target. After that, the full calibration is comparably easy. The values for telescope azimuth and elevation, as well as the position of the GRAVITY K-Mirror and HWP, are given in the header of each file (see \autoref{tab:header}). With this information, the Mueller matrix, as given in \autoref{equ:fullmmatrix}, can be calculated. To use the presented calibration model for polarized observations with GRAVITY, we put all the information into a small python package \textsc{VLTIpol}, which is publicly available\footnote{\url{https://github.com/widmannf/VLTIpol}}. The package includes tools to read out all the necessary information from a file header and calculate the Mueller matrix of a specific observation from these observations. This makes the polarimetric observing mode easy to calibrate and fully available to the community.

\section{Differential effects}\label{sec:differential}
Apart from the absolute calibration of instrumental polarization, another important question is whether the differential birefringence between the telescopes causes errors in astrometric measurements. This topic was initially addressed by \cite{Lazareff2014}. However, in their analysis, they used small random perturbations of the telescopes to estimate the phase error and get the best alignment of the optical components in GRAVITY. We can now extend their analysis, as we do not have to work with random perturbations but have the measurements of the instrumental polarization for the four UTs.

\subsection{Conventions - Jones formalism}\label{sec:jones}
For the analysis of differential effects between the telescopes, we need a description of the propagated phase, which is not possible with the Stokes formalism. For this case, we will instead use the Jones formalism, which ultimately allows us to estimate the phase errors introduced by differential birefringence. In this formalism, the state of polarization of an electric field is described by a complex Jones vector:

\begin{equation}
    j = \left(\begin{array}{c} 
                E_x \\ 
                E_y
        \end{array}\right)
        = \left(\begin{array}{c} 
                A_x \cdot e^{i\phi_x}\\ 
                A_y \cdot e^{i\phi_y}
        \end{array}\right).
\end{equation}
A change in radiation is again described by a matrix, the 2x2 complex Jones matrix:
\begin{equation}
    j_{out}=J\cdot j_{in},
\label{equ:jones_prop}
\end{equation}
with the input Jones vector $j_{in}$ and the output vector  $j_{out}$. One of the main disadvantages of the Jones formalism is that Jones vectors always represent fully polarized light.
To be able to deal with partially polarized light, one has to use the Hermitian coherence matrix of the electric field \citep[see][]{Born1999}:
\begin{equation}
    C = \left(\begin{array}{cc} 
                \langle E_xE_x^* \rangle & \langle E_xE_y^* \rangle \\ 
                \langle E_yE_x^* \rangle & \langle E_yE_y^* \rangle \\ 
        \end{array}\right),
\label{equ:coh_polarization}
\end{equation}
with $E^*$ being the complex conjugate of the field. The coherence matrix has real values on the diagonal elements, corresponding to the total intensity in the x and y directions. The trace of the matrix gives the full intensity of the field:
\begin{equation}
    I = \mathrm{Tr}(C) = \langle E_xE_x^* \rangle + \langle E_yE_y^* \rangle .
\end{equation}
The off-diagonal elements are complex and describe the correlation between the x and y components of the electric field.

The degree of polarization of an electric field described by a coherence matrix is \citep{Born1999, Gil2004}:
\begin{equation}
    \mbox{DOP} = \sqrt{\frac{2\cdot \mathrm{Tr}(C^2)}{\mathrm{Tr}(C)^2}-1} .
\label{equ:coh_dop}
\end{equation}
Applying \autoref{equ:jones_prop} to the electric field, one can see that the Jones matrix can be applied to the coherence matrix in the following way \citep{Hamaker2000}:
\begin{equation}
    C_{out} = J\cdot C_{in}\cdot J^*.
\end{equation}

The coherence matrix as described by \cite{Born1999} was then used to describe the effects of polarization in an interferometer by \cite{Hamaker1996, Hamaker2000, Smirnov2011}. For the combinations of two telescopes of an interferometer, $m$ and $n$, the coherence matrix can be written as follows:
\begin{equation}
    C_{m,n} = \left(\begin{array}{cc} 
    \langle E_{m,x}E_{n,x}^* \rangle & \langle E_{m,x}E_{n,y}^* \rangle \\ 
    \langle E_{m,y}E_{n,x}^* \rangle & \langle E_{m,y}E_{n,y}^* \rangle \\ 
     \end{array}\right).
\end{equation}

The response of an interferometer to an electromagnetic signal \vec{E} depends on the polarization properties of the two telescopes in the baseline. When one describes the polarization properties with Jones matrices $J_m$ and $J_n$ the coherence matrix propagates as follows:
\begin{equation}
\begin{split}
    V_{m,n} &= 2~\langle J_m\cdot E_m\left(J_n\cdot E_n\right)^H\rangle \\
    &= 2~J_m \cdot \left(\begin{array}{cc} 
        \langle E_{m,x}E_{n,x}^* \rangle & \langle E_{m,x}E_{n,y}^* \rangle \\ 
        \langle E_{m,y}E_{n,x}^* \rangle & \langle E_{m,y}E_{n,y}^* \rangle \\ 
        \end{array}\right) \cdot J_n^H \\
    & = 2~ J_m \cdot C^{in}_{m,n} \cdot J_n^H
\end{split}
\label{equ:rime}
\end{equation}
Here $^H$ indicates the hermitian matrix. This approach of describing the influence of polarization effects on the interfering electric fields is taken from the radio interferometer measurement equation (RIME), which was introduced by \cite{Hamaker1996}. This quantity $V_{m,n}$ in the RIME formalism has different names across the literature. Defined as \textit{coherency matrix} by \cite{Hamaker2000}, \textit{visibility matrix} by \cite{Smirnov2011} or \textit{cross-coherence matrix} in \cite{Lazareff2014}, we will adopt \textit{visibility matrix} from \cite{Smirnov2011}, as the matrix indeed contains the complex visibilities \citep[as outlined by][]{Smirnov2011}. The visibility matrix describes the correlation of two electric fields $E_i'$, which have been modified by the properties described by the Jones matrix: $E_i' = J_i\cdot E_i$ and the individual elements describe the correlation in two individual feeds, in our case for an orthonormal $xy$-basis. 

As shown by \cite{Smirnov2011} one can extend the RIME formalism to include further instrumental effects such as bandwidth smearing or can take an arbitrary brightness distribution into account, which will ultimately lead to a formulation of the Van-Cittert Zernike theorem out of the RIME formalism. However, in our case, we want to use this approach to describe the effect that two non-ideal telescopes have on the interferometric signal and do not include any further effects. We will, therefore, only use the formalism as given in \autoref{equ:rime}. Even more so, we can directly take into account that GRAVITY can only measure two linear polarization states at a time. The two polarization states are split by a Wollaston prism, which results in an angle of \SI{90}{\degree} between the two states. When we align our coordinate system with one of the two states, we can directly use the x and y components of \autoref{equ:rime}. The useful quantities are then $V_{m,n}^{x,x}$ and $V_{m,n}^{y,y}$.

As indicated before, these two quantities describe the complex visibility at the detector level of a (potentially polarized) point source, with only the polarization effects in the two light paths taken into account. From the complex visibility, we can then calculate further useful quantities, such as the correlated flux:
\begin{equation}
F_{m,n}^x = \left|V_{m,n}^{x,x}\right|,
\end{equation}
and the total photometric flux, given by the sum of the flux arriving at the individual telescopes:
\begin{equation}
I_{m,n}^x = \frac{1}{2} \left(V_{m,m}^{x,x} + V_{n,n}^{x,x}\right).
\end{equation}
With these two quantities, we can then calculate the maximum visibility or fringe contrast, which is just the fraction of flux that is coherent:
\begin{equation}
\nu_{m,n}^x = |V_{m,n}^x| / I_{m,n}^x.
\label{equ:coh_matrix}
\end{equation}
As we are at this point ignoring all effects except the instrumental polarization and birefringence in the two light paths, this quantity gives the loss in coherence due to the polarization effects. When the two light paths are identical $\nu_{m,n}$ equals one, and it reduces when we lose coherence due to differential polarimetric effects. We will look at the coherence loss in the VLTI in \autoref{sec:diff_contrast}.

Similarly, we can also calculate the fringe phase:
\begin{equation}
\Phi_{m,n}^x = Arg(V_{m,n}^{x,x}).
\end{equation}
$\Phi_{m,n}$ equals zero for identical light paths and deviates from zero otherwise. A deviation from zero fringe phase will be corrected by the fringe tracker and will not impact the coherence. However, when the science and fringe tracking target have a different polarization state, they will also have a different fringe phase. This will add an error to the measured science phase as it is referenced to the fringe tracking phase. The phase error for typical GRAVITY targets is studied in  \autoref{sec:diff_phase}.

The equations shown here for the $x$ component are the same for the $y$ component. With this concept, we have everything in hand to calculate the interferometric response to polarized targets, taking into account the polarization properties of the individual telescopes. We will continue fitting the VLTI model to get the Jones matrix of each telescope and then study the differential effects and their impact on observations.

\subsection{Fitting individual telescopes}
Similar to the approach in \autoref{sec:modeling}, we again fit the model to the calibration data. But this time, we treat each telescope individually. The main concept of the model does not change: we reuse the model as described in \autoref{equ:modelall}. In this case, however, we exchange the Mueller matrix with a Jones matrix. Using Fresnel calculus, the Jones matrix that describes the linear retardation, as well as the reflection, can be written as follows: \begin{equation}
J = \frac{r_p}{2}\left( 
\begin{array}{cc}
r_s \cdot \exp(i\delta) & 0 \\
0 & r_p \\
\end{array}
\right).
\label{equ:jones_mod}
\end{equation}
All our measurements are given as normalized Stokes vectors. One can convert them into Jones vectors with the following formula:
\begin{equation}
    j = \frac{1}{\sqrt{2}}\left(\begin{array}{c} 
                \sqrt{1+Q} \\ 
                \sqrt{1-Q} \cdot \exp\left(- i\arg (U+iV)\right)
        \end{array}\right)
        = \left(\begin{array}{c} 
                A_x \\ 
                A_y \cdot e^{i\delta}
        \end{array}\right).
\end{equation}
As we do not have the full phase information in such a vector, we have to allow for an additional phase for each input state. In the end, we fit
\begin{equation}
    j_{out} = J \cdot \left(e^{i\phi} \cdot j_{in}\right),
\end{equation}
where $\phi$ is an arbitrary phase factor that we ignore. For the Jones description of a mirror, we must go back to three parameters per mirror group instead of the two parameters from \autoref{equ:mueller_mod}. So, for each of the three mirror groups, we individually fit a reflection coefficient in the $p$ and $s$ directions and a phase difference between the two reflections. Furthermore, we split our data set into individual telescopes, which divides the amount of available data for each fit by four. While the fit is very similar to the previous one in \autoref{sec:calibration}, we need to refit here as we now treat the different telescopes individually.

\begin{figure}
	\centering
	\includegraphics[width=0.45\textwidth]{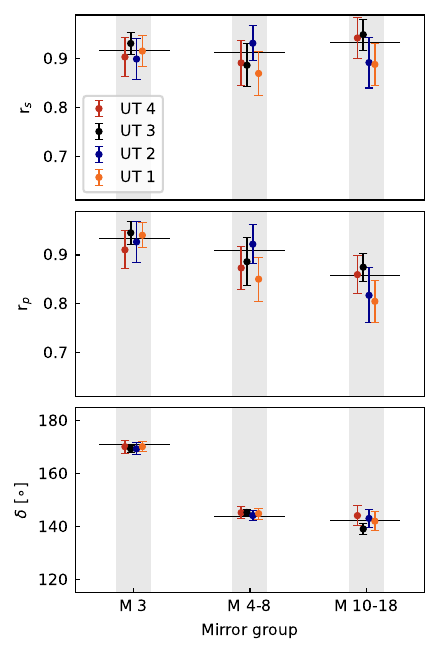}
	\caption{Comparison of the fitting values for the individual telescopes. The three plots show the reflectivity in the s and p directions, as well as the phase difference for each group of mirrors. The data points of all telescopes are grouped for one mirror group, and the fitting value for the combined fit is shown as a black horizontal line. $r_s$, $r_p$ and $\delta$ are defined in \autoref{equ:jones_mod}.}
	\label{fig:tel_comp}
\end{figure}

The first result for the individual fits is shown in \autoref{fig:tel_comp}, where the reflectivity in s and p direction, as well as the phase difference between s and p, is shown. From this first look, we can conclude that in general, all the values are similar. This confirms the initial assumption that the birefringence in the individual telescopes is on the same order. Furthermore, the newly fitted values are close to the ones from the combined fit in \autoref{sec:calibration}. We see, however, some scatter in the reflection coefficients of the different telescopes. This means we have some degree of differential attenuation in the VLTI, which leads to some loss in fringe contrast. The phase difference, however, is very stable over all telescopes. This is a good sign, as it indicates that we have very little differential retardance, which would show up as a phase error in the observations. We will look into both effects in more detail in the following.

\subsection{Fringe contrast}\label{sec:diff_contrast}
\begin{figure}
	\centering
	\includegraphics[width=0.45\textwidth]{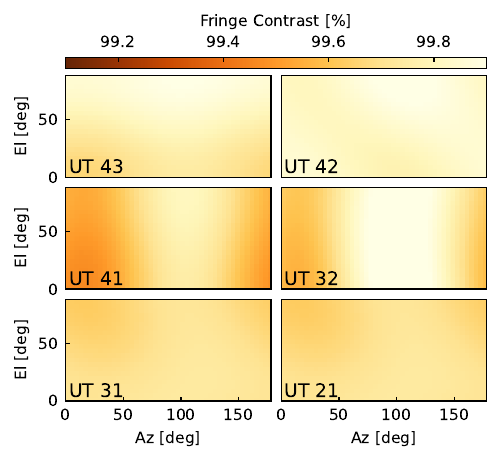}
	\caption{Fringe contrast in the six baselines shown for all possible telescope positions}
	\label{fig:ev_fringeloss}
\end{figure}
With the response of each individual telescope, we can ask the question of how much the different instrumental polarization in each light path influences the interferometric observations. For this, we use the matrix representation of the visibilities (\autoref{sec:jones}). First, we look at the loss in fringe contrast. We assume that our fringe-tracking object is unpolarized. Nevertheless, instrumental polarization may introduce a small degree of polarization in the incoming light. As shown in \autoref{equ:coh_matrix} one can calculate the fringe contrast by the quotient of correlated and total flux, basically asking the question of how much of the total incoming light interferes. As the instrumental polarization depends on the telescope's position, we calculate the fringe contrast for a grid of telescope positions. This grid reaches from 0 to 90 \si{\degree} in elevation and 0 to 180 \si{\degree} in azimuth. In azimuth, the telescopes can rotate between 0 and 360 \si{\degree}, but the polarization signal repeats after \SI{180}{\degree}, so it is sufficient to calculate the values in this range.

The fringe contrast for all telescope positions is shown in \autoref{fig:ev_fringeloss}. For all baselines, the fringe loss is always well below \SI{1}{\percent}. This means we have a fringe contrast of above \SI{99}{\percent} for unpolarized fringe tracking targets. As the average on-sky fringe contrast for bright calibrators is on the order of \SI{90}{\percent}, we can conclude that the fringe loss due to polarization is insignificant.

\subsection{Eigenvectors}
Following \cite{Lazareff2014}, we use polarization eigenvectors to describe the polarization properties of the mirror train. A polarization eigenvector is defined as a linear input polarization that results in a linear output polarization. \cite{Lazareff2014} have shown that each Jones matrix describing the VLTI has two distinct eigenvectors. They are generally not orthogonal to each other but close to orthogonal. With the now-measured Jones matrices for each telescope, we confirm both findings. To minimize the phase error \cite{Lazareff2014} suggested that one should align one of the polarization directions, measured on the detector, orthogonal to one of the output eigenvectors. 

The effect of phase errors due to the differential birefringence depends on the direction of the measured polarization on the detector in the instrument. We will, therefore, use GRAVITY as an example for now, but the findings apply to all instruments.

As discussed in \autoref{sec:gravity}, the propagation of the polarization direction through GRAVITY is calibrated with the help of Fibered Polarization Rotators. We can, therefore, assume that the two detector polarizations correspond to the same directions in the VLTI lab. For the proposed alignment of the eigenvector with the polarization axis on the detector, one could use the HWP. In the current instrument setup, the HWP rotates with the K-Mirror (see \autoref{sec:gravity}), which does not achieve this alignment. In the following, we will look into the phase errors due to the non-alignment of the eigenvectors and the detector and estimate how much they could be reduced with the optimal alignment.

\subsection{Phase errors}\label{sec:diff_phase}
\begin{figure*}
	\centering
	\includegraphics[width=0.9\textwidth]{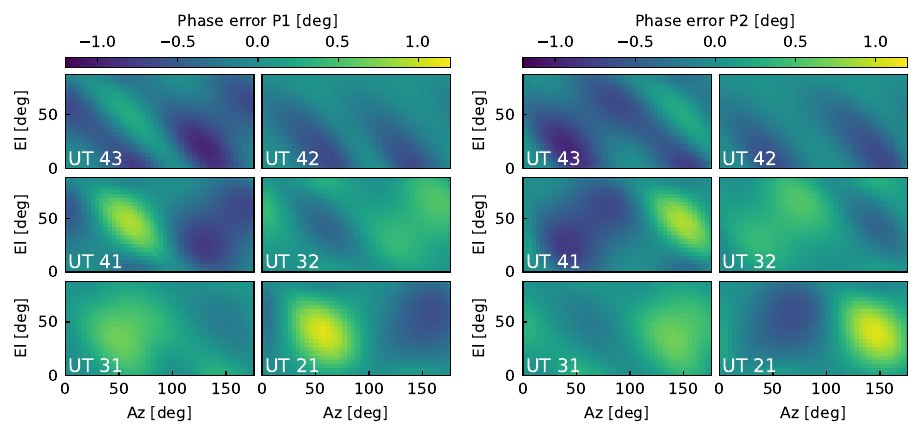}
	\caption{Error in the visibility phases due to differential birefringence for all telescope positions. The left two columns show the error for the first polarization P1 and the right two for the second polarization P2.}
	\label{fig:ev_phaseerr}
\end{figure*}
\begin{figure*}
	\centering
	\includegraphics[width=0.9\textwidth]{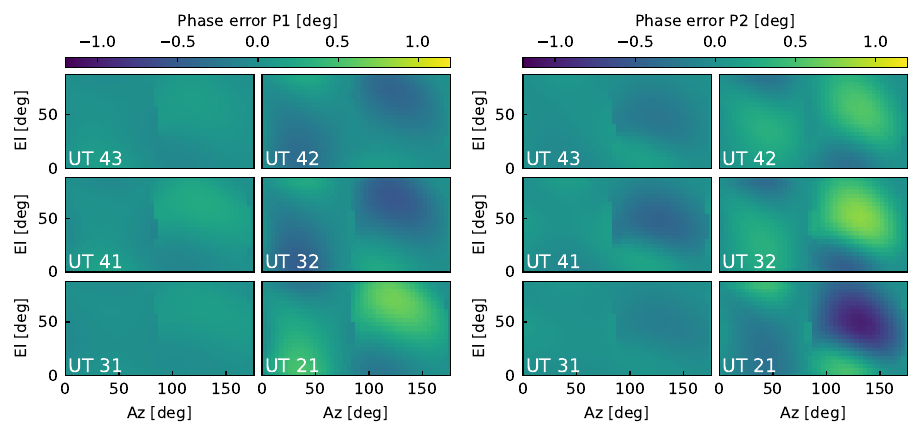}
	\caption{Same representation as in \autoref{fig:ev_phaseerr}, but this time the phase error if the GRAVITY half wave plates track the eigenvector of the individual telescopes.}
	\label{fig:ev_phaseerr_cor}
\end{figure*}

The measured visibility phase is referenced to the fringe-tracking object. This can be the same as the science target or a nearby star. As discussed in \cite{Lazareff2014}, this can lead to a measurement error if the fringe-tracking and the science target have a different polarization state. In the case of two different targets, the fringe-tracking object will most likely be a star and, therefore, unpolarized. In some cases, the fringe-tracking source could be slightly polarized due to foreground dust \citep[see for example][]{Buchholz2013} or intrinsic polarisation of, for example, a dusty giant \citep{Haubois2019}. However, this should only be a few percent and is irrelevant here as the science target can have a much higher polarization. In the case of an (almost) unpolarized fringe-tracking target and a highly polarized science target, the polarimetric response of the mirror train is different for the two targets. This different instrumental polarization will introduce a phase error, as the science phases are referenced to the fringe-tracking phases.

For the following tests, we assumed a science target with a linear polarization of \SI{30}{\percent}. This value is chosen, as it is a likely value for the Galactic Center supermassive black hole Sgr~A* in its flaring state \citep[see e.g.][]{Genzel2010}, which is one of the most extreme levels of NIR polarization known in celestial bodies. The following coherence matrix represents such a polarization state (see \autoref{equ:coh_polarization}):
\begin{equation}
    C = \left(\begin{array}{cc} 
                0.65 & 0 \\ 
                0 & 0.35 \\ 
        \end{array}\right).
\end{equation}
As discussed in \autoref{sec:jones}, the intensity of the electric field is given by the diagonal elements, and the degree of polarization is calculated following \autoref{equ:coh_dop}, which gives a \SI{30}{\percent} polarization degree for this matrix. Following \autoref{equ:coh_matrix} the measured phase of a target is just the argument of the visibility matrix. We calculate this for the unpolarized fringe-tracking object and the slightly polarized science object and subtract the two phases from each other to take the phase referencing into account. As for the fringe contrast, we again calculate this for each baseline and each telescope position and show the results in \autoref{fig:ev_phaseerr}. Theoretically, there is another degree of freedom, which is the orientation of the polarization vector on sky. This is given by the intrinsic polarization of the source as well as the parallactic angle. However, as this is just a rotation, it is redundant with the telescope azimuth and would shift the pattern in \autoref{fig:ev_phaseerr} to the left or right. We, therefore, ignore this for now. 

As shown in \autoref{fig:ev_phaseerr}, there is a small phase error that depends on the telescope position. On average, the error is of the order of \SI{0.3}{\degree}, with maximal values of \SI{1.1}{\degree}. However, the two polarizations show a somehow opposite pattern. If one averages the two polarizations, as one would probably do it for astrometry measurements anyway, to increase the SNR, the phase error reduces. For the average value, the mean phase error is \SI{0.2}{\degree}, with a maximum value of \SI{0.8}{\degree}. In the simplest case, a single point source, the phases relate to the position on sky with the following formula:
\begin{equation}
    \Phi = 2\pi \vec{s}\cdot\vec{B} / \lambda,
\end{equation}
where $\Phi$ is the measured phase, $\vec{s}$ the measured position on sky and $\vec{B}$ the baseline length. Inverting this formula and using a baseline length of \SI{100}{\meter} and a wavelength of \SI{2.2}{\micro\meter} a phase error of \SI{0.8}{\degree} corresponds to an astrometric error of \SI{10}{\micro as}.

It was already shown by \cite{Lazareff2014} that this error can be improved if the output eigenvector of the telescope is aligned with the axis of the polarization measurement on the detector. We can confirm these previous results that each telescope always has two eigenvectors, which are roughly, but not exactly, \SI{90}{\degree} apart. If one aligns one detector polarization with one eigenvector, the astrometric error of this measurement drops to 0. However, the second polarization still shows a significant phase error. From simulating all the different options, we found that the lowest overall phase error can be achieved if one uses the average of the two eigenvector angles:
\begin{equation}
\Bar{\phi}_{EV} = \frac{1}{2}\left(\phi_{EV1} + \phi_{EV2}-\textstyle\frac{\pi}{2}\right).
\label{equ:ev_ev}
\end{equation}
We can align the detector polarization with this vector by rotating the half-wave plate (HWP) by this angle. If we do so for each telescope individually, we reach the phase errors as shown in \autoref{fig:ev_phaseerr_cor}. One can see that the phase error has been reduced in comparison without the HWP rotation in \autoref{fig:ev_phaseerr}. The mean phase error in this case is significantly decreased to \SI{0.1}{\degree} with maximum values up to \SI{0.9}{\degree}. Again, we can further improve this by averaging the two polarizations to mean values below \SI{0.1}{\degree} and a maximum error of \SI{0.7}{\degree}. We, therefore, see that while such an alignment improves the situation, it is only a small improvement if we work with the mean phase. In \autoref{fig:ev_phaseerr_cor}, the most significant values can be seen in all baselines with UT3. This is due to the fact that for UT3 the two eigenvectors are less orthogonal than for the other telescopes. Averaging the two eigenvector angles adds a slightly higher phase error than for the other baselines. 

To reach the smallest possible phase error, as shown in \autoref{fig:ev_phaseerr_cor}, one would need to track the eigenvector with the HWP. This tracking angle is a pure telescope property and, therefore, does not depend on the polarization on sky or the parallactic angle. It could be implemented as a look-up table based on the derived values. The angles are very similar for the four telescopes but not identical. This means one must align each telescope individually to get the smallest phase error. This might not be desired as then each telescope would have a different orientation of the polarization axis on the detector. Furthermore, tracking of the HWP could introduce systematic effects into the metrology measurement. We find that the phase error without tracking is comparably small and does not dominate over other systematic effects. 

In conclusion, one does not need to align the detector with the eigenvectors as long as one uses the mean phase for the highest resolution astrometry but the individual polarizations for the polarization measurement and the imaging.

\section{Conclusion}\label{sec:conclusion}
In this paper, we have presented the first complete polarization study of the VLTI and GRAVITY. As expected, both the observatory as well as the instrument itself show polarization effects. In both cases, we characterized the effects and built up a calibration model to calibrate for the instrumental effects in observations. We have outlined the necessary steps in observation, data reduction, and calibration to execute polarimetric measurements with GRAVITY. The capabilities were then demonstrated by remeasuring the polarization properties of the Galactic Center star IRS~16C, which is in excellent agreement with the literature. 

We have also shown that differential birefringence between the light paths of the VLTI UTs is not a dominant error source, as the four light paths were constructed with great care to minimize differential birefringence. For a typical observation of a calibrated source, our studies have shown a phase error due to differential birefringence of below \SI{1}{\degree}. Even for the extreme case with a \SI{30}{\percent} polarized science target, this results in only around \SI{10}{\micro as} astrometric error. This error can be further reduced when using the average of the two polarizations for astrometry and the individual signal for polarimetry. The fringe contrast in such a case is only reduced by around \SI{1}{\percent}. 

We, therefore, demonstrated that observations with GRAVITY do not suffer from strong effects due to birefringence and that GRAVITY can be used for polarimetric observations. This can be done in very different ways. With IRS~16C we showed that GRAVITY can measure the polarization quantities of even slightly polarized targets with very good precision. One can also study the temporal evolution of polarized targets, as it was done in \cite{GRAVITY2018, Gravity2020pol} for a bright Sgr~A* flare. Furthermore, the polarization information can not only be extracted for the full intensity, but also for the measured intensity in each spectral channel. One can, therefore, even map polarization changes over the spectral range of GRAVITY. As all of this comes together with the unprecedented resolution of GRAVITY, this opens up a wide new range of possibilities to do polarimetry in the near-infrared. 

\begin{acknowledgements}
We are very grateful to our funding agencies (MPG, ERC, CNRS [PNCG, PNGRAM], DFG, BMBF, Paris Observatory [CS, PhyFOG], Observatoire des Sciences de l’Univers de Grenoble, and the Fundação para a Ciência e Tecnologia), to ESO and the Paranal staff, and to the many scientific and technical staff members in our institutions, who helped to make GRAVITY a reality. F. W. has received funding from the European Union’s Horizon 2020 research and innovation programme under grant agreement No 101004719. A.A., and P.G. were supported by Fundação para a Ciência e a Tecnologia, with grants reference SFRH/BSAB/142940/2018, UIDB/00099/2020 and PTDC/FIS-AST/7002/2020.
\end{acknowledgements}
 
\bibliography{bibfile} 

\begin{appendix}
\section{Modelling of the coated mirrors}\label{app:protected}
\begin{figure}
	\centering
    \includegraphics[width=0.45\textwidth]{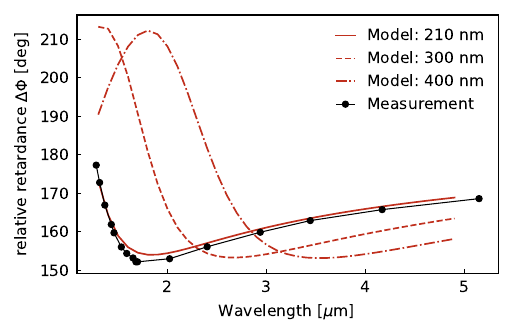}
	\caption{Measured and modeled phase difference for the protected silver mirrors. The black points show the measured data, and the solid red line shows the model with the thickness that best fits the data. For illustration, two more models with higher thicknesses are shown in dashed and dotted lines.}
	\label{fig:coating}
\end{figure}
For the coated mirrors, we have to modify the Fresnel equations introduced in \autoref{sec:modeling} to account for an additional thin layer on top of the bare metal mirror. This is usually done with techniques from Elipsometry, a method to determine the thickness of a thin film by measuring its effect on the polarization of a reflected beam. A more complete overview of this topic can, for example, be found in \cite{Jellison1999} or in Chapter 29 of \cite{Goldstein2003}.

For this analysis, we start with the equations \ref{equ:fresnels} and \ref{equ:fresnelp}, which give the reflectance for a single layer. Now, we cannot assume that one of the refractive indices is equal to one, but we have to assume that both indices can be arbitrary. By using Snell's law we can modify equations \ref{equ:fresnels} and \ref{equ:fresnelp} to the following:
\begin{equation}
R_s = \frac{n_2\cos(\Theta_1) - n_1\cos(\Theta_2)}{n_2\cos(\Theta_1) + n_1\cos(\Theta_2)},
\end{equation}
\begin{equation}
R_p = \frac{n_1\cos(\Theta_1) - n_2\cos(\Theta_2)}{n_1\cos(\Theta_1) + n_2\cos(\Theta_2)},
\end{equation}
We then need to evaluate these equations twice, once for the transition from air to the thin film ($R_{1,s}$ and $R_{1,p}$) and once for the transition from the thin film to the base layer ($R_{2,s}$ and $R_{2,p}$). With those four values, we can then get the reflectance for the full mirror with the thin coating:
\begin{equation}
R_{f,s} = \frac{R_{1,s}+R_{2,s}\cdot\exp{\left(-ib\right)}}{1+R_{1,s}\cdot R_{2,s}\cdot\exp{\left(-ib\right)}},
\end{equation}
\begin{equation}
R_{f,p} = \frac{R_{1,p}+R_{2,p}\cdot\exp{\left(-ib\right)}}{1+R_{1,p}\cdot R_{2,p}\cdot\exp{\left(-ib\right)}},
\end{equation}
with the factor b:
\begin{equation}
b=\frac{4\pi d n_f\cos\Theta_f}{\lambda}
\end{equation}
with the thickness of the layer $d$, the refractive index of the layer $n_f$, and the angle within the layer $\Theta_f$, which is calculated with Snell's law from the incidence angle: $\sin\Theta_i = n_f\sin\Theta_f$. The quantities of the diattenuation and the phase difference are then calculated as before. 

We use the thin film formulas for all silver mirrors in the VLTI, as they have a protective coating. We have measurements for the phase difference of one of the mirrors available (F. Delplancke, private communication), but we do not know specifics about the coating itself. To model the mirrors, we assume a protective coating out of $Al_2O_3$, a reasonably typical coating for silver mirrors. To estimate the thickness of the coating, we fit the calculated phase difference $\Delta = \arg R_{f,p} - \arg R_{f,s}$ to the measured data and fit for the thickness. We find that a thickness of \SI{210}{\nano\meter} represents the data reasonably well, as shown in \autoref{fig:coating}. We adopt this value for all silver mirrors in the initial model, but it is unclear if it is a good representation for all mirrors. However, as the diattenuation and retardance are fitted to data in a later step, it should not influence the calibration model.

\section{Wavelength dependency of the model}\label{app:model}
\begin{figure}
	\centering
    \includegraphics[width=0.45\textwidth]{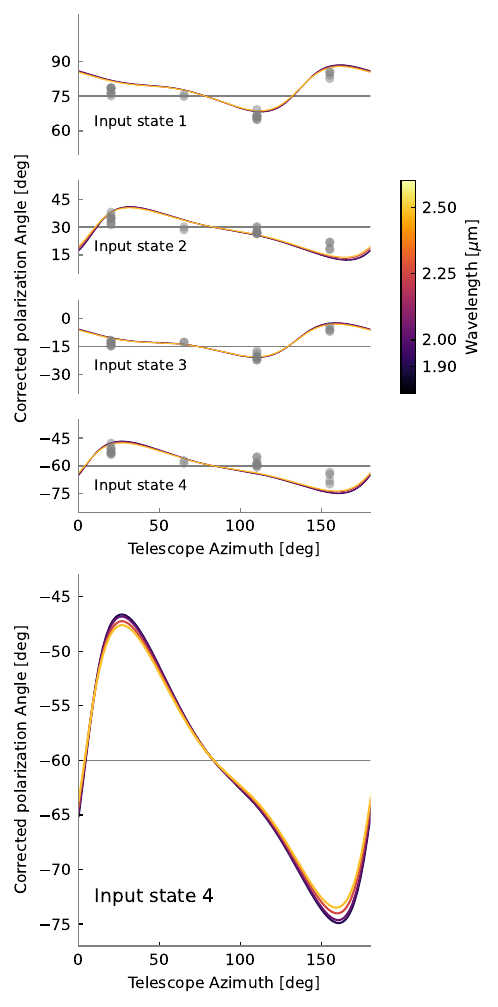}
	\caption{Effect of different wavelengths on the polarization model of the VLTI lightpath. The top panel is similar to \autoref{fig:pol_ang}, showing the measured polarization angle (corrected for field rotation) for all four telescopes and all linear input states in grey. The model is shown in four wavelengths (1900, 2000, 2250, and \SI{2500}{nm}). As the model repeats after \SI{180}{\degree}, only the range from 0 to 180 degrees is shown. The bottom panel shows a scaled-up view of the model for state 4 without the data.}
	\label{fig:wavelength}
\end{figure}
The data we used to calibrate our model was taken with a laser at a wavelength of \SI{1908}{nm}. This is slightly lower than the science wavelength of GRAVITY, which is between 2000 and \SI{2500}{nm}. As the refractive index of the mirrors we model changes with wavelength, the difference in wavelength introduces a small error into the model. However, for the used materials, the change in refractive index over the wavelength range is $\pm (0.1 + 2i)$. This is comparably small and should not introduce a big error. To test the effect, we modeled the VLTI lightpath at four different wavelengths: at \SI{1908}{nm}, where the laser is, and at the minimum, mean, and maximum wavelength of the GRAVITY spectrum (2000, 2250, and \SI{2500}{nm}). The model at different wavelengths is shown in \autoref{fig:wavelength}. The difference between the four models is small, with a mean absolute difference of 0.3 between the laser wavelength and the mean science wavelength (\SI{2250}{nm}) and a maximum difference of below \SI{1}{\degree}, depending on the telescope position. This is in the same order as the repeatability of the experimental input states and smaller than the uncertainty of the fitted phase shifts (\autoref{sec:calibration}). We conclude that the wavelength of the laser is not a dominant error in our model. Especially for longer observations, it will average out, and for shorter observations, it will add an uncertainty of below one degree, dependent on the telescope position.

\section{Mueller matrix of GRAVITY}\label{app:matrices}
The measured Mueller matrices for GRAVITY are:\\
For the K-Mirror:
\begin{equation}
M_{KM} = \left( 
\begin{array}{cccc}
1 & -0.0146 & 0 & 0 \\
-0.0146 & 1 & 0 & 0 \\
0 & 0 & -0.8949 & -0.4461 \\
0 & 0 & -0.4461 & -0.8949 
\end{array}
\right).
\end{equation}
For the HWP, we assume a perfect matrix:
\begin{equation}
M_{HWP} = \left( 
\begin{array}{cccc}
1 & 0 & 0 & 0 \\
0 & 1 & 0 & 0 \\
0 & 0 & -1 & 0 \\
0 & 0 & 0 & -1 
\end{array}
\right).
\end{equation}
For the remaining optics in off-axis mode:
\begin{equation}
M_{off-axis} = \left( 
\begin{array}{cccc}
1       &  0.0036 &  0.0250 &  0.0110 \\
-0.0237 & -0.7143 & -0.6685 & -0.1674 \\
0.0261  & -0.6743 &  0.7041 &  0.1918 \\
0.0294  &  0.4122 &  0.2723 & -0.8742
\end{array}
\right).
\end{equation}
and in on-axis mode:
\begin{equation}
M_{on-axis} = \left( 
\begin{array}{cccc}
1       &  0.0612 &  0.0496 &  0.0141 \\
-0.0901 & -0.7457 & -0.7160 & -0.0836 \\
 0.0870 & -0.6884 &  0.7768 &  0.1066 \\
-0.0246 &  0.3469 & -0.0438 & -0.8325
\end{array}
\right).
\end{equation}

\end{appendix}

\end{document}